\documentclass[epj,nopacs,final]{svjour}
\usepackage{graphics}
\usepackage{epsfig}
\usepackage{multirow}
\begin{document}
%

\title{What is the $\mathbf{W b \bar b}$, $\mathbf{Z b \bar b}$ or $\mathbf{t \bar t b
  \bar b}$  irreducible background to the light Higgs boson searches  at LHC?}

\titlerunning{What is the $\mathbf{W b \bar b}$, $\mathbf{Z b \bar b}$ or $\mathbf{t \bar t b
  \bar b}$  irreducible background \ldots}
	
\author{Borut Paul Kersevan\inst{1}\fnmsep\inst{2} \and El\. zbieta Richter-W\c{a}s\inst{3}\fnmsep\inst{4}\fnmsep\thanks{ Supported in part by  Polish Government grant KBN 2P03B11819.}}

\authorrunning{B. P. Kersevan \and E. Richter-W\c{a}s}

\institute{Faculty of Mathematics and Physics, University of Ljubljana,
 Jadranska 19, SI-1000 Ljubljana, Slovenia. \and
 Jozef Stefan Institute, Jamova 39, SI-1000 Ljubljana, Slovenia. \and
 Institute of Computer Science, Jagellonian University, 30-072 Krakow,
 ul. Nawojki 11, Poland.\and
 Institute of Nuclear Physics, 30-055 Krakow, ul. Kawiory 26a, Poland.}

\date{Received: date / Revised version: date}

\abstract{The $ W  b \bar b$, $ Z  b \bar b$ and $t \bar t b \bar b$ production at
LHC are irreducible backgrounds for possible observability of the Standard Model  and
Minimal Supersymmetric Standard Model light Higgs boson in processes involving
associated $WH$, $ZH$ and $t \bar t H$ production followed by the $H \to b \bar b$ decay.
The comparison presented in this paper uses the background estimates obtained with (a) complete massive
matrix element implemented in {\tt AcerMC} Monte Carlo generator and (b) {\tt PYTHIA}
implementation of the inclusive $W$, $Z$, and $t \bar t$ production, followed by the
parton showering mechanism.  Both approaches lead to a production of the final state of
interest but differ in the approximations used.  The focal point of this study is 
the comparison of the two approaches when estimating the background contributions 
to the light Higgs boson searches at the LHC.}

\maketitle

\section{Introduction}

The $ W b \bar b$, $ Z b \bar b$ and $t \bar t b \bar b$ production at LHC has been
recognized , see e.g. \cite{ActaB31,ActaB30}, to present the most substantial irreducible
backgrounds for the Standard Model (SM) and Minimal Supersymmetric Standard Model (MSSM)
light Higgs boson observability in the associated production, namely $WH$, $ZH$ and $t
\bar tH$, followed by the $H \to b \bar b$ decay. The 'light Higgs boson' is in this context 
understood as the Higgs boson having the mass between 90~GeV and 130~GeV, thus describing
the SM and MSSM Higgs boson(s) in the mass range indicated by the excess observed in
searches of LEP experiments \cite{LEP1,LEP2}.

The potential of the ATLAS detector at LHC for the SM and MSSM Higgs boson observability
in the $t \bar t H$ production has already been carefully studied and documen\-ted in
\cite{ActaB30}  and \cite{ATL-PHYS-TDR}.
The proposed analysis requires four identified (tagged) b-jets, reconstruction of both
top-quarks decaying in hadronic and leptonic modes and an observable peak in the invariant
mass distribution of the remaining b-jet pair. The irreducible $t \bar t b \bar b$
background is estimated to contribute about 60-70\% to the total background, which
consists mostly of the processes with a $t \bar t$ pair participating in the final
state. The expected significance is deemed to be of the order of 3.6~$\sigma$ for the
Higgs-boson mass of 120~GeV at integrated luminosity of $30 fb^{-1}$, with the expected
signal-to-background ratio in the mass window rated to be 32\%. The total contribution
from the $Wjjjjjj$ backgrounds was estimated to be an order of magnitude smaller than the
one consisting of $t \bar t jj$ events and the contribution of the $t \bar t Z$ production
process was estimated to be negligible.

The potential for the Higgs boson observability in the $WH$ $(W \to \ell \nu)$ production, 
con\-sidered as rather weak, is documen\-ted in \cite{ActaB31} and \cite{ATL-PHYS-TDR}.
The expected production rates would be sufficient for the signal observability in the
mass range around 120~GeV only if the backgrounds from $Wb \bar b$ and $t
\bar t$ events could be efficiently suppressed. For this channel both b-quarks are
required to be tagged as b-jets and the reconstruction of a peak in the invariant mass of
the b-jet system concentrated in the interval $\pm~20$~GeV around the expected Higgs mass
could lead to an evidence of the signal. The identification of the accompanying lepton is
also required in order to trigger the data acquisition.  Relatively simple topology of the
final state does not leave much room for a severe optimisation of kinematic cuts. The only
possibility which can be explored seems to be using a veto on an additional jet to
suppress reducible backgrounds or angular correlations between reconstructed b-jets
themselves and/or between b-jets and leptons. Within the low luminosity operation, the
irreducible $W b \bar b$ background contributes about 35\% of the total background to this
channel, the expected signal-to-background ratio in the given mass window being on the
order of a few percent. The expected significance is estimated to be on the level of
3.0~$\sigma$ for the Higgs-boson mass of 120~GeV and integrated luminosity of $30 fb^{-1}$.

Much less promising, if not hopeless to consider, is the observability of the $ZH$
production with subsequent leptonic Z-boson decay. Nevertheless, for the sake of
completeness, such a study was documented in \cite{ActaB31}.  Both b-quarks are required
to be tagged as b-jets and the accompanying leptons are required to be identified in order
to trigger the data acquisition as well as for a reconstruction of the resonant peak
around the Z-boson mass.  The presence of a peak in the reconstructed invariant mass of
the b-jet system concentrated in a mass window of $\pm~20$~GeV around the expected Higgs
mass would lead to an evidence of the signal. The irreducible $Z b \bar b$ background
contributes about 80\% of the total background to this channel, the expected
signal-to-background ratio in the mass window being on the order of a few percent. The
expected significance is estimated to be about 1.0~$\sigma$ for the Higgs-boson mass of
120~GeV and integrated luminosity of $30 fb^{-1}$.

It is thus evident that good theoretical understanding of the irreducible $t \bar t b \bar
b$, $W b \bar b$ and $Z b \bar b$ backgrounds is crucial for the light Higgs boson
observation at LHC.

In the presented study the two available, albeit qualitatively different, approaches for
simulating these backgrounds are compared.  The first one (ME) is to use the lowest order
massive matrix elements of the {\tt AcerMC} generator \cite{AcerMC} which lead to the
required final state. The latter is subsequently completed with the initial and final
state radiation simulated via parton showering as implemented in the {\tt PYTHIA}
\cite{Pythia62} or {\tt HERWIG} \cite{Herwig63} generators.  The second one (PS) is to
simulate inclusive $t \bar t$, $W$ and $Z$ production using the native {\tt PYTHIA} or
{\tt HERWIG} implementations and subsequently obtain accompanying b-quarks using the
parton shower approximation only. Both approaches have their caveats. Using the complete
$2 \to 4$ matrix element might not be sufficient as the b-quarks could appear at the
different steps in the evolution of the partonic cascade and not necessarily only at the
hard process level. Resorting to the parton shower approach on the other hand tends in
several cases to underestimate hardness of the radiated partons and does not reproduce
well their topological configurations.

In the presented comparison the approach of concentrating not on the partonic
distributions but on the reconstructed experimental quantities, i.e. jets and isolated
leptons, is chosen. The generated events are thus treated with a simplified reconstruction
procedure using the algorithms of the fast simulation of the ATLAS detector at LHC
\cite{ATL-PHYS-98-131} and subsequent fiducial cuts on reconstructed jets and isolated
leptons are applied roughly as foreseen for this type of physics at LHC detectors:
geometrical acceptance for b-jets and isolated leptons identification down to
pseudorapidity $\eta$ of 2.5 and transverse momenta threshold for jets and leptons of
15~GeV.  Although leptons will be identified with the lower thresholds, the 15~GeV
transverse momenta represents roughly what is needed for triggering on such events. This
limits the given comparison to the topological configurations close to those which will be
selected by the experimental analysis. In jet reconstruction a simple procedure in form of
a cone algorithm is performed, using a cone radius of 0.4 in the pseudorapidity -
azimuthal angle plane. Subsequently, the procedures of jet energy calibration and
b-tagging are applied. In case jet-veto was also applied, events with additional jets having
transverse momenta $p_T>30$~GeV and $|\eta|<5.0$ were rejected.  Efficiencies for b-tagging
and lepton identification are not included in the given numbers, only the efficiencies for
jet  and b-jet reconstruction are taken into account. More details on the
performance of the applied algorithms can be found in \cite{ATL-PHYS-98-131}.

The {\tt AcerMC} Monte Carlo generator code and its interfaces to {\tt PYTHIA} generator
were used in the given evaluation. The generated statistics was typically about $10^6$
events for the ME and $10^8$ events for the PS simulation chain. The ME events were
generated with {\tt AcerMC} matrix element implementations and the PS events were
generated with the default settings of {\tt PYTHIA 6.2}. The ME events were further
completed with the initial and final state radiation to assure more realistic jet
reconstruction efficiencies and multiplicities, thus leading to a better description of
the interaction kinematics.  The CTEQ5L\cite{cteq5l} parton density functions were used for all estimations
and proton-proton collision at 14~TeV centre-of-mass energy were simulated. A similar
study could also be repeated using {\tt HERWIG} instead of {\tt PYTHIA} generator.

 At this point
it might be relevant to mention that for the inclusive $W$ and $Z/\gamma^*$ production the so
called  improved parton shower algorithm is implemented in {\tt PYTHIA}, i.e. some
higher-order corrections are integrated; as shown in \cite{Sjostrand1998}, it gives good
description of the complete $p_T^W$ spectrum. A corresponding improvement is also expected
for the production of jets in association with the $W$ boson.  For the $t \bar t$
production the standard parton algorithm is still used {\tt PYTHIA}.

The comparisons between these two generation approaches for $W b \bar b$, $Z b \bar b$ and $t
\bar t b \bar b$ events are discussed in Section 2, 3 and 4 of the paper. Final conclusions
are summarised in Section 5.

\section{The $\mathbf{Wb \bar b}$ irreducible background}

In this section  the irreducible $W b\bar b$ background to the Higgs
boson searches in the $WH$ production, followed by the $H \to b \bar b$ decay, is
discussed. The evaluation is based on two simulation approaches, ME and PS, as specified
below:

\begin{itemize}
\item
 {\bf ME:} Use the $2 \to 4$ matrix element for $q \bar q \to W(\to \ell \nu) g^*(\to b \bar
 b)$ process as implemented in \cite{AcerMC}. The $\sigma \times BR$ = 33.2~pb for single
 leptonic flavour W-boson decay.  The $g^* \to b \bar b$ splitting is coded already at the
 level of the matrix element. This matrix element represents the lowest order contribution
 to the $\ell b \bar b$ final state. The initial (ISR) and final state radiation (FSR) are
 simulated with parton shower of {\tt PYTHIA}, followed by hadronisation in order to
 complete the event generation.
\item
 {\bf PS:} Use the $2 \to 1 $ matrix element for $q \bar q \to W $ process as implemented in
 {\tt PYTHIA}, followed by the ISR/FSR and hadronisation. The $\sigma \times BR$ = 17200
 pb for the single flavour leptonic W-boson decay. Gluon splitting in the ISR partonic
 cascade is the source of b-quarks in the event. The implementation of this process
 includes ISR/FSR modeled with {\it an improved parton shower approach} 
\cite{Sjostrand1998,Sjostrand2000}
 to match/merge with higher order matrix element calculations for the
 $p_T^W$ spectra.
\end{itemize}

Considering the heavy flavour content of the cascade, a part of the higher order
corrections is {\it a priori} already included in the parton shower approach
(c.f. \cite{Hautmann,Puljak}) whereas only the lowest order term for the $g^* \to b \bar
b$ splitting is present in the matrix element calculations.

However, as the $2 \to 4$ matrix element with activated ISR and FSR is used, a part of the
higher-order corrections is also to some extend included (e.g. additional branchings of $b
\to b g$ are made possible), however without a rigorous check on the possible {\it
double-counting} \cite{Catani}. In order to limit the double counting possibility the
scale of the ISR/FSR shower is tuned to avoid the radiated partons being harder than the
hard-process ones. What is not included in the $2 \to 4$ matrix element calculations is
the contribution from events where gluon and quark interact in the hard process to produce
the $W$-boson, with the gluon splitting into b-quarks occurring in the further steps of
the cascade. As well not included is the contribution where the final state gluon in the
hard process decays into light quarks, while b-quarks appear in another branch of the
partonic shower.

\begin{figure}
\begin{center}
     \epsfig{file=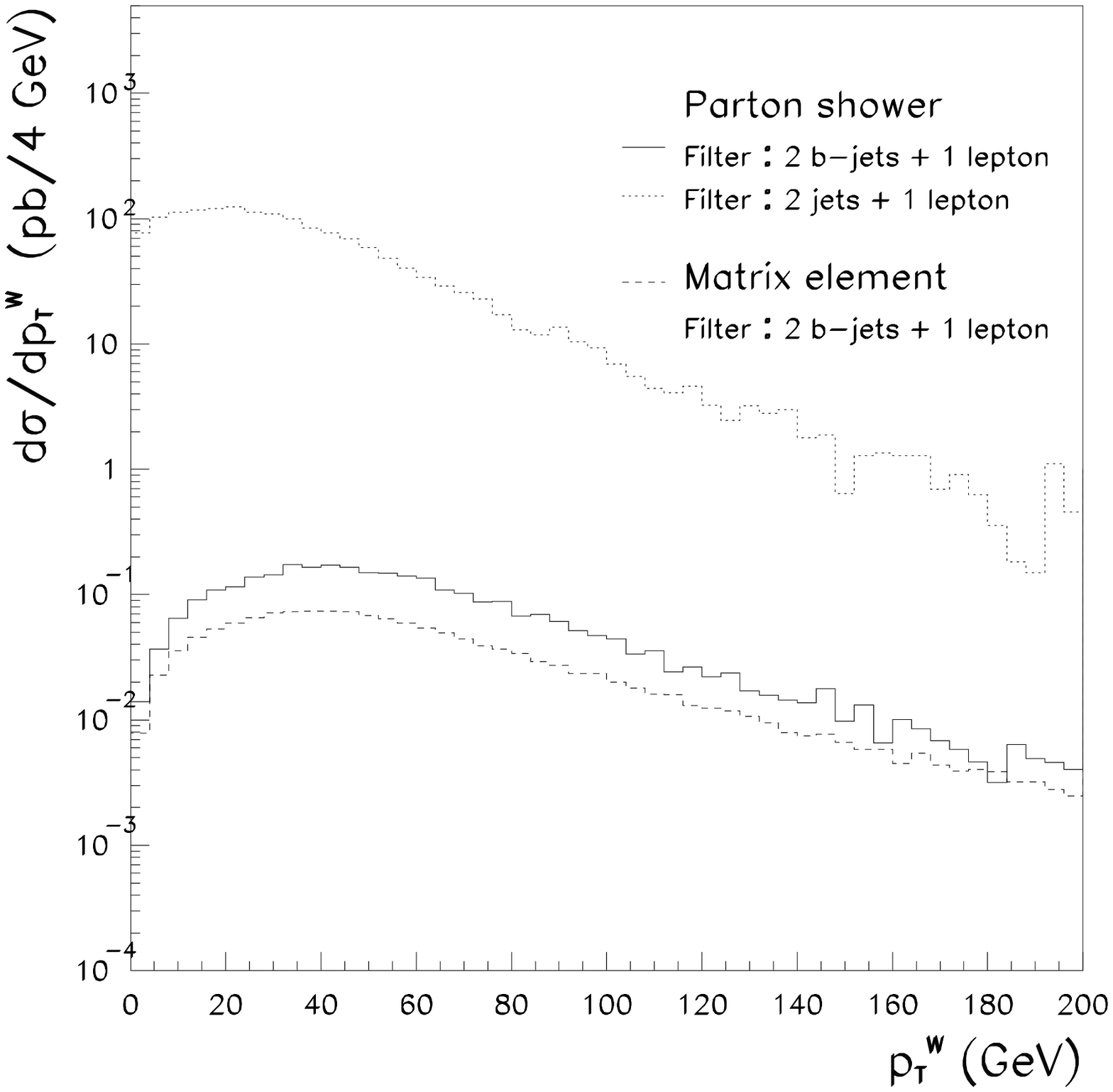,width=4.5cm}\\
     \epsfig{file=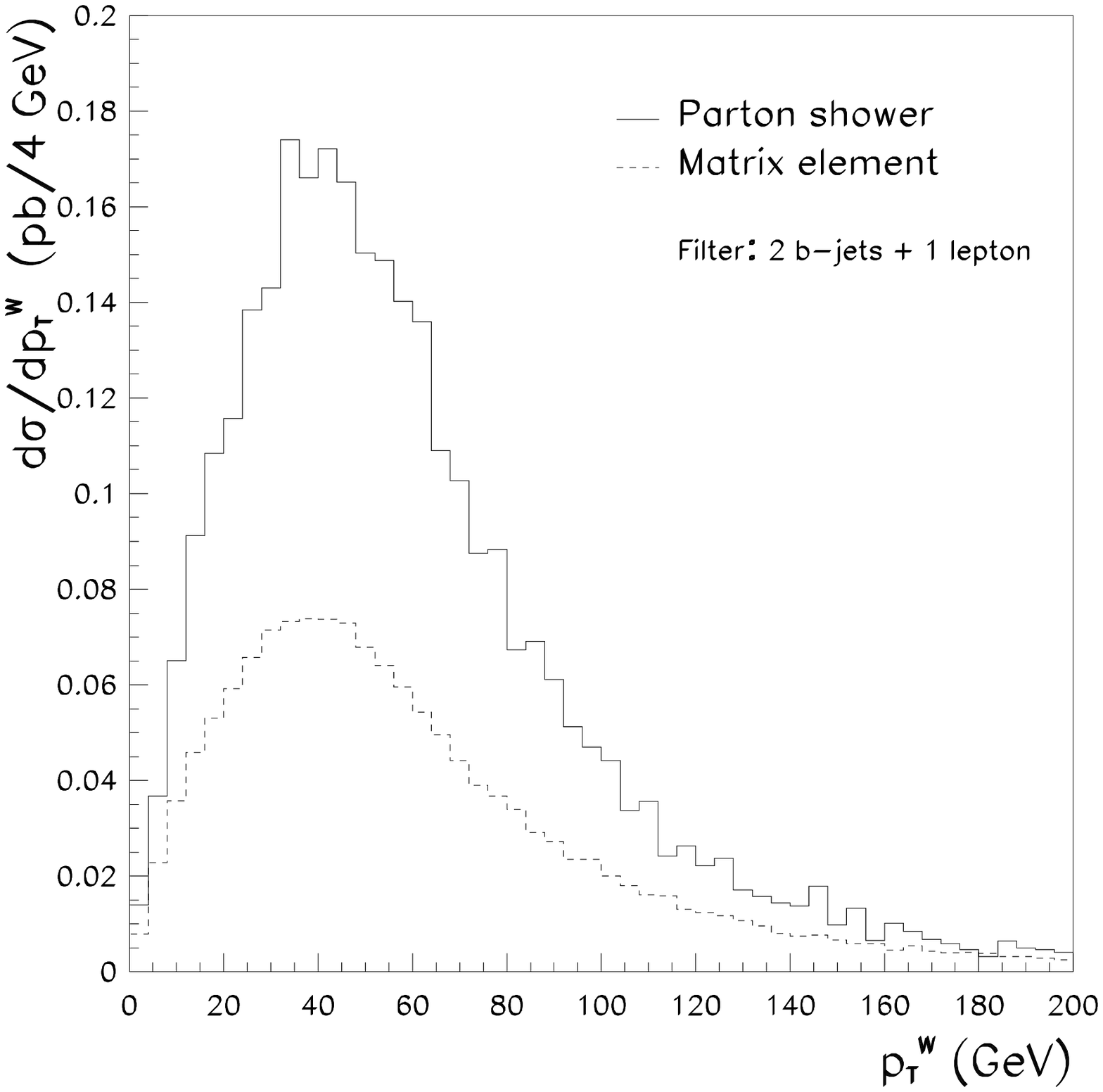,width=4.5cm}\hskip -0.3cm
     \epsfig{file=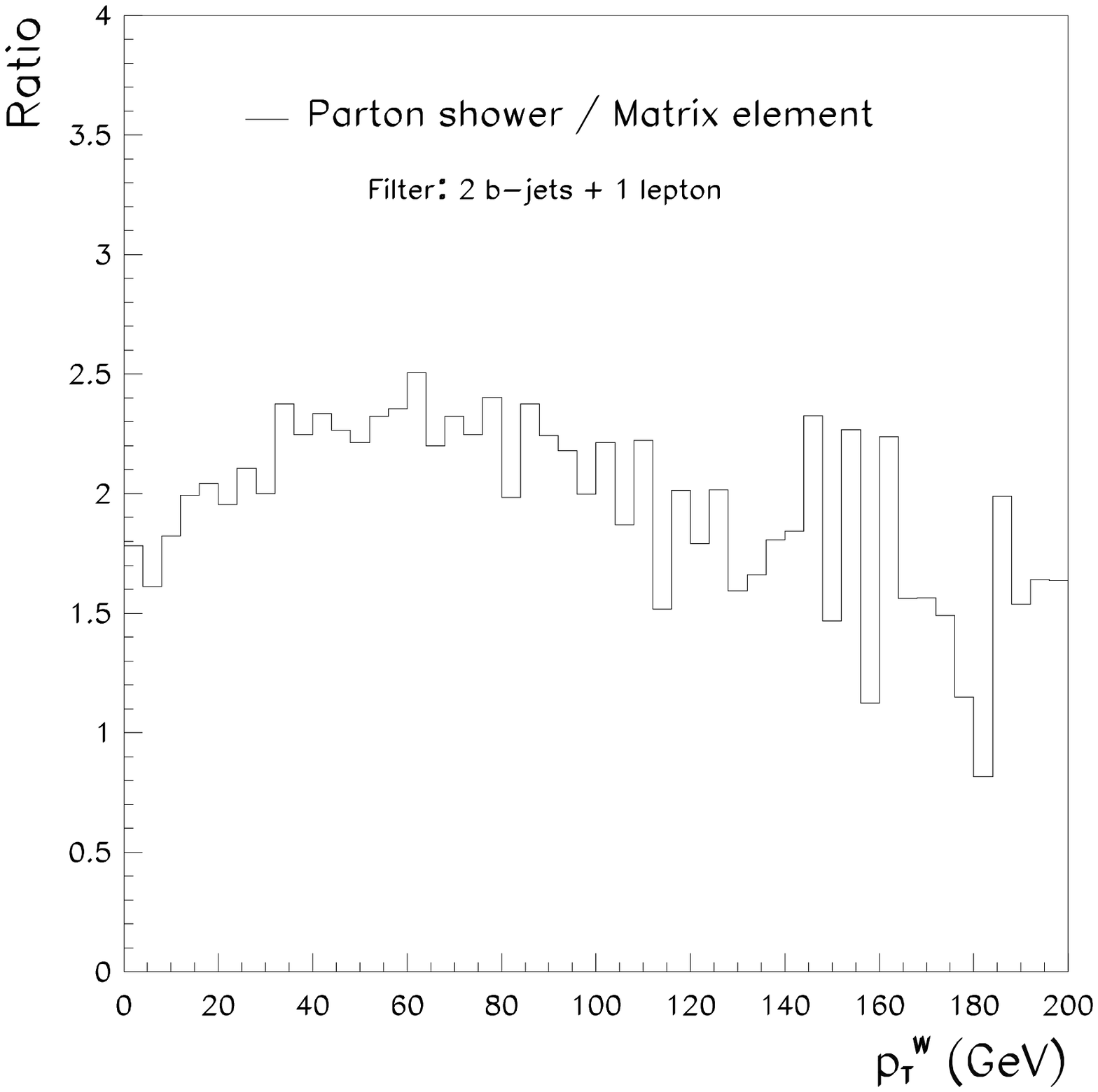,width=4.5cm}
\end{center}
\caption{\em
The  transverse momenta of the W-boson.
Solid line denotes the PS events, dashed the ME events. Events were filtered as
specified in the figure.
\label{FS2:a}} 
\end{figure}

Comparison of the differential distributions of the $p_T^W$ spectra is presented in
Fig.~\ref{FS2:a}. In the top plot the PS distribution after filtering on events with
reconstructed 2 jets + 1 lepton is shown, along with PS and ME distributions after
filtering on exactly 2 b-jets + 1 lepton.  In the lower range of the $p_T^W$ spectra the
PS events have different slope when requiring 2 jets or 2 b-jets; requiring 2 b-jets
strongly suppresses the selection of events in the low $p_T^W$ range. After filtering on 2
b-jets the slope and normalisation of the PS and ME events agree relatively well, the
normalisation ratio being on the level of 1.5-2.5.  A closer look (bottom plots) indicates
a substantial enhancement of PS events in the range $p_T^W =
40-120$~GeV. For much higher $p_T^W$ the
ME predictions start to exceed the PS ones.

\begin{table} 
\newcommand{\lstrut}{{$\strut\atop\strut$}}
  \caption {\em Cross-section for the $q \bar q \to \ell \nu b \bar b$ ME and $q \bar q
  \to W $ production PS with $W \to \ell \nu$ decay (single flavour).  Efficiencies for
  b-tagging and lepton identification are not included, only included are the efficiencies
  for jet  reconstruction and b-jet tagging.
\label{TS2:a}} 
\vspace{2mm}  
\begin{center}
\begin{tabular}{lll}
\hline\noalign{\smallskip}
Selection &  $q \bar q \to \ell \nu b \bar b $ & $q \bar q \to W (\to
          \ell \nu)$ \\
          & ME   &  PS \\
\noalign{\smallskip}\hline\noalign{\smallskip}
Generated  $\sigma \times BR$  & 33.2 pb  & 17200 pb    \\
\noalign{\smallskip}\hline\noalign{\smallskip}
Two b-jets + one lepton  &  1.46 pb  & 3.10 pb  \\
$m_{\rm bb-jets} = 100-140$ GeV & 0.16  pb  & 0.23  pb  \\
\noalign{\smallskip}\hline\noalign{\smallskip}
\parbox[c]{3.8cm}{Two b-jets + one lepton + jet-veto}  & 1.13  pb  & 1.55 pb  \\
\noalign{\smallskip}
$m_{\rm bb-jets} = 100-140$ GeV & 0.12  pb  & 0.12  pb  \\
\noalign{\smallskip}\hline
\end{tabular}
\end{center}
\end{table}

Table~\ref{TS2:a} quantifies the expected cross-sections for the inclusive production
after requiring a reconstructed $\ell b b$ final state. The PS predictions turn out to be
50-100\% higher than the ME ones but are still quite compatible for events with the
invariant mass of the b-jet system in the range of interest. After requiring jet-veto,
important for selection of the $WH$ channel to suppress the $t \bar t$ background, the ME
predictions agree with the PS ones in the mass range of interest.  The numbers in this table
illustrate that the slope of the invariant mass distribution of the b-jet system and jet
multiplicities is quite different for different simulation approaches.

For the sake of the evaluation consistency, factorisation/renormalisation energy scale
$Q^2=m_W^2$ was used throughout the ME and PS event generation.  The factorisation/renormalisation
scale dependence for this process is of the order of 20\% at most, which
was estimated by using a range of different definitions of the energy scale implemented in
\cite{AcerMC}.

 The fact that the overall normalisations differ, one can
naively interpret as an effect indicating that the {\it effective} cascade branching into
heavy flavour quarks is more intense in PS events, quantitatively by almost a factor of 1.5-2.5
higher. The feature was discussed in more detail in \cite{ATL-COM-PHYS-2001-032},
the conclusions being that it rather reflects a sum of several effects, not just a simple
enhancement in the effective cascade branching.

\begin{figure}
\begin{center}
     \epsfig{file=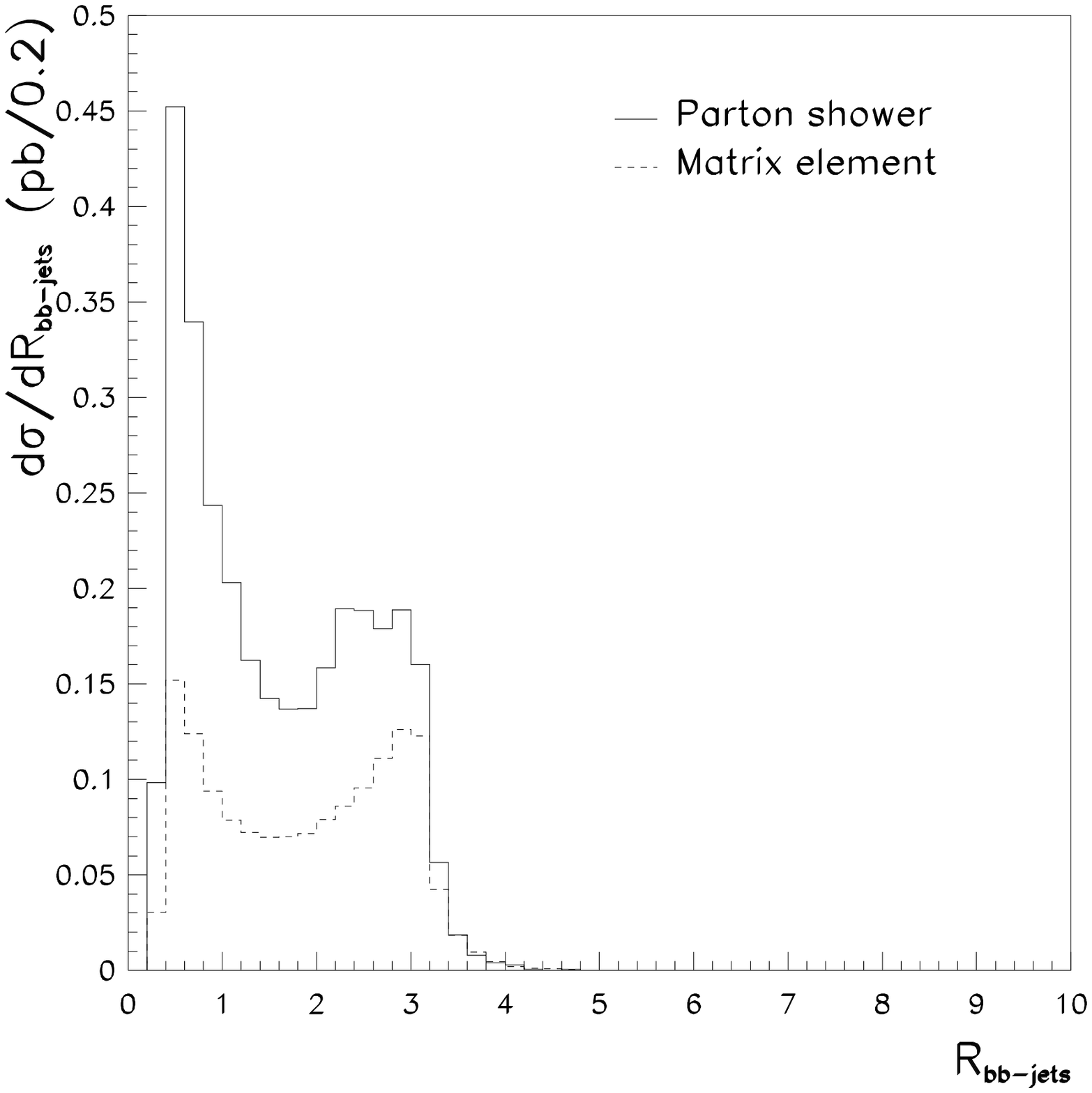,width=4.5cm}\hskip -0.3cm
     \epsfig{file=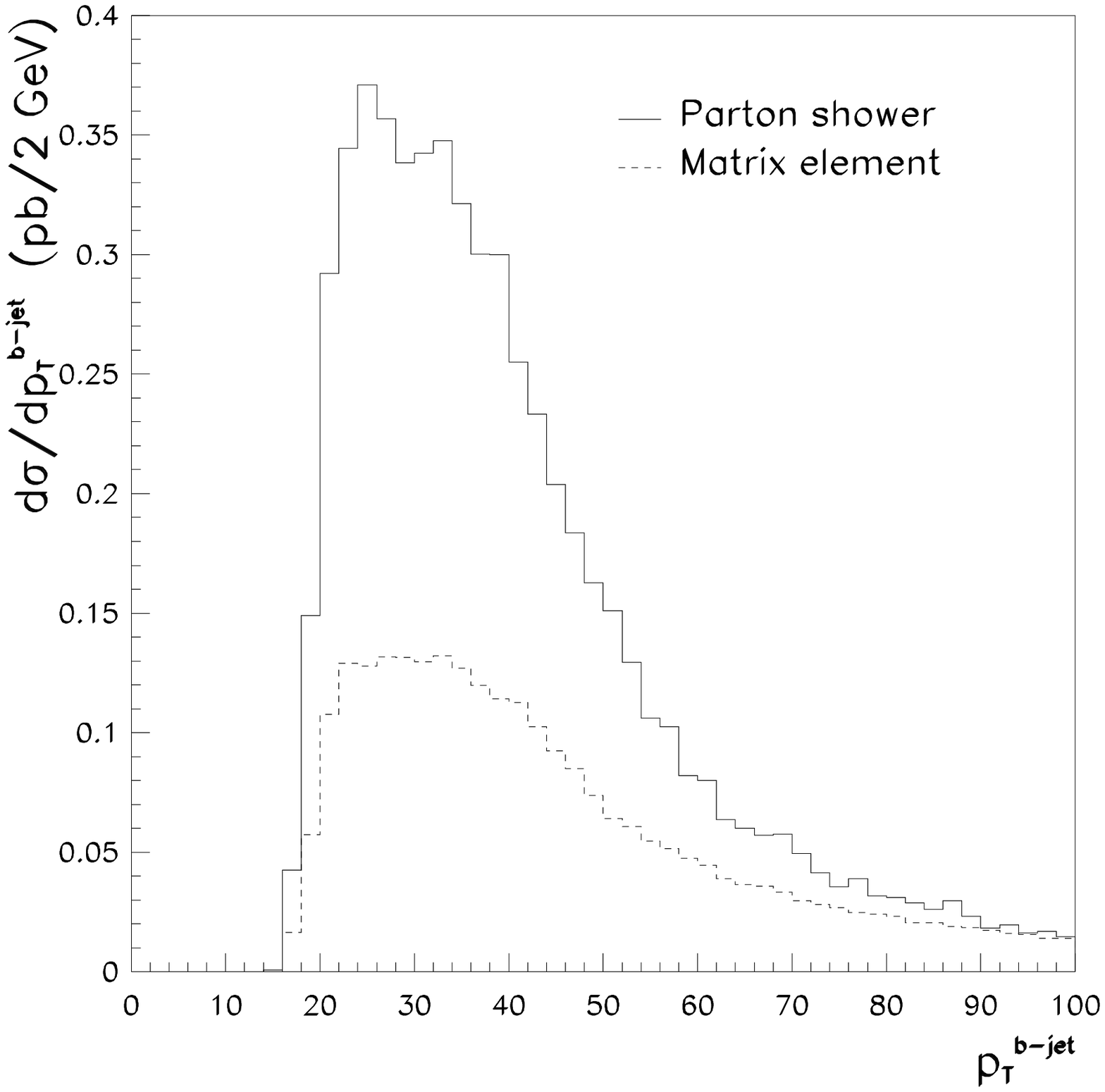,width=4.5cm}\\
     \epsfig{file=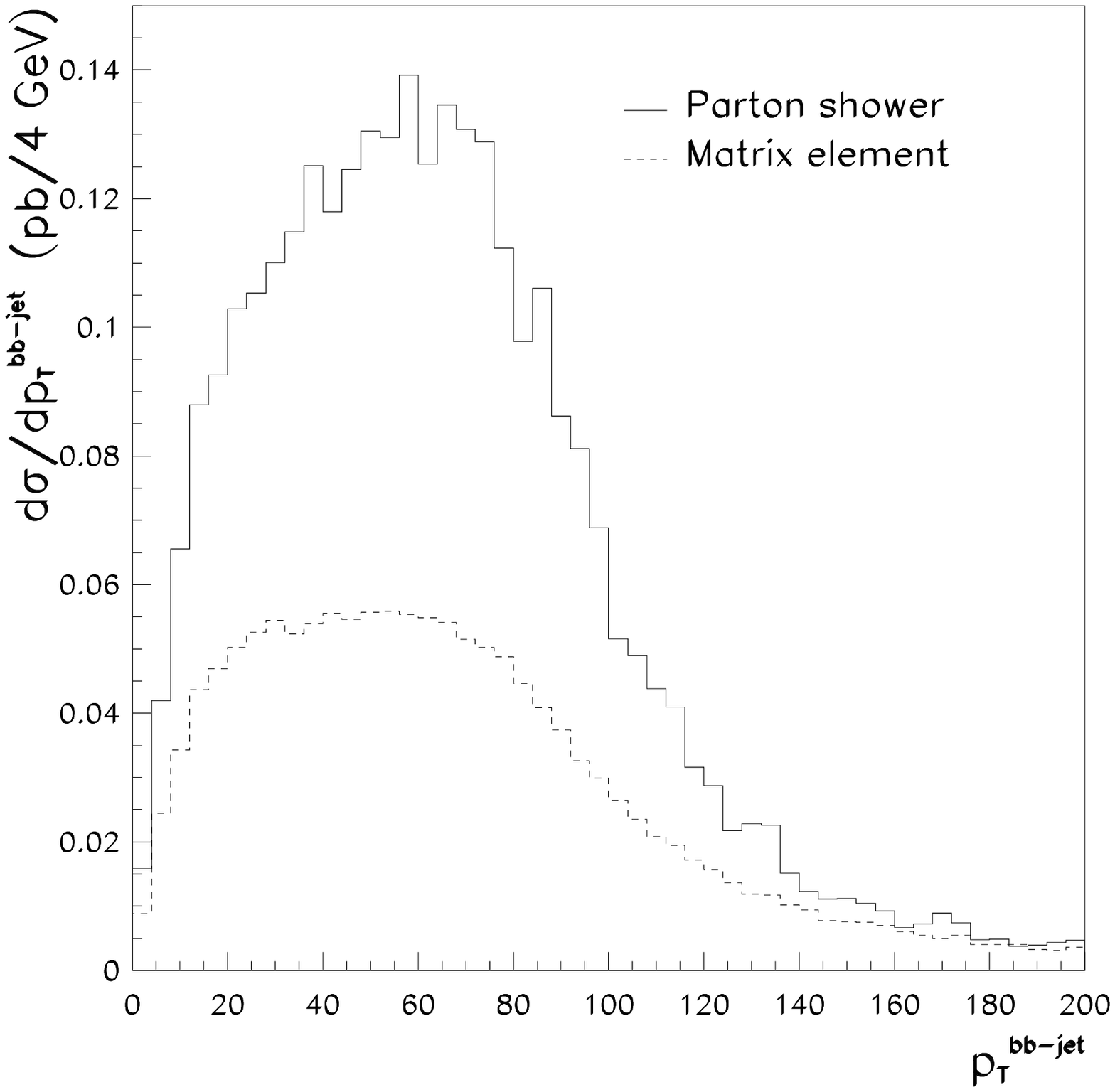,width=4.5cm}\hskip -0.3cm
     \epsfig{file=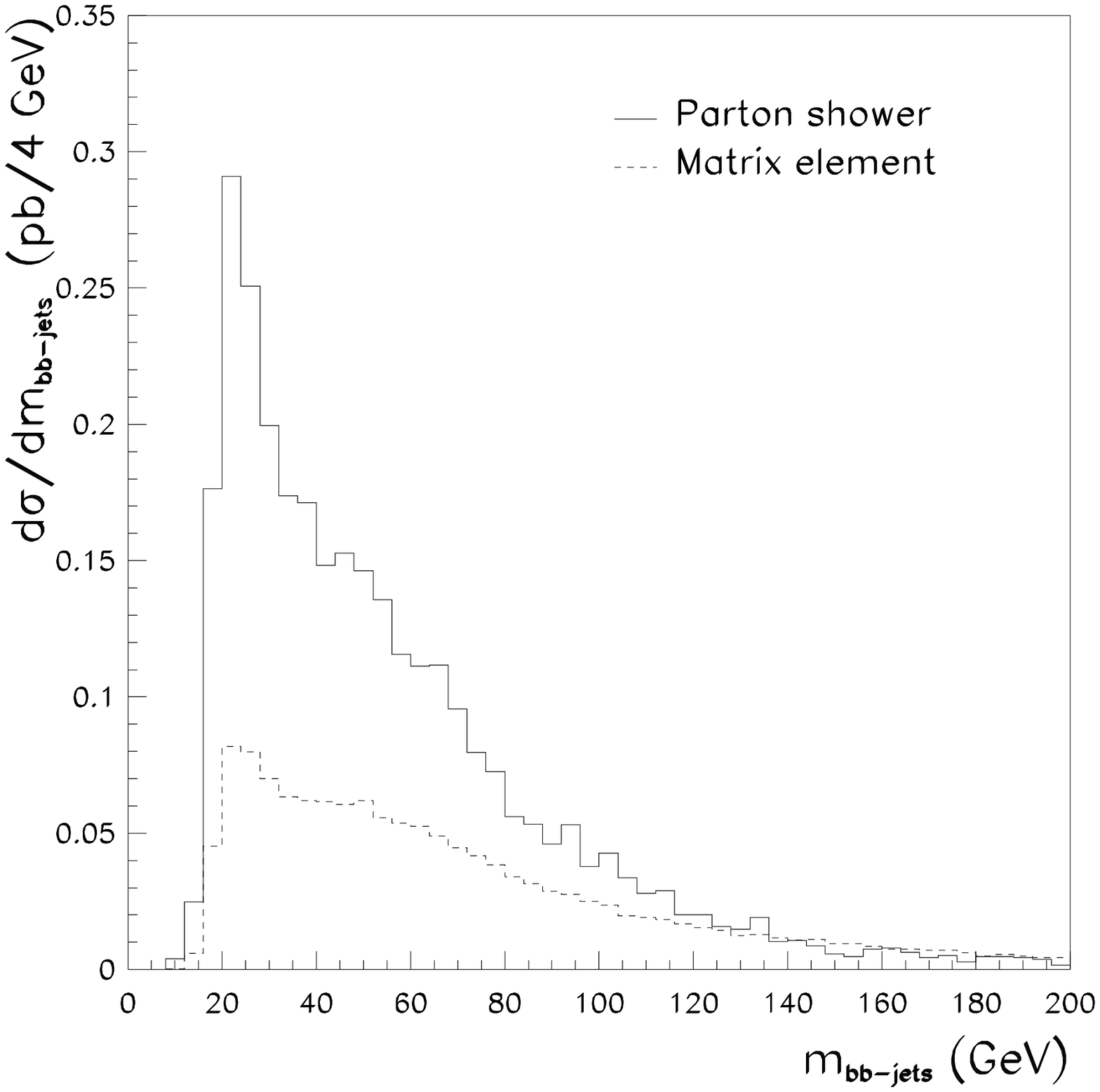,width=4.5cm}

\end{center}
\caption{\em
Top: the cone separation between b-jets, transverse momenta of the individual b-jets;
Bottom: transverse momenta of the b-jet system and the invariant mass distribution of the
b-jet system.  Solid line denotes the PS events, the dashed one ME events.
\label{FS2:b}} 
\end{figure}

\begin{figure}
\begin{center}
     \epsfig{file=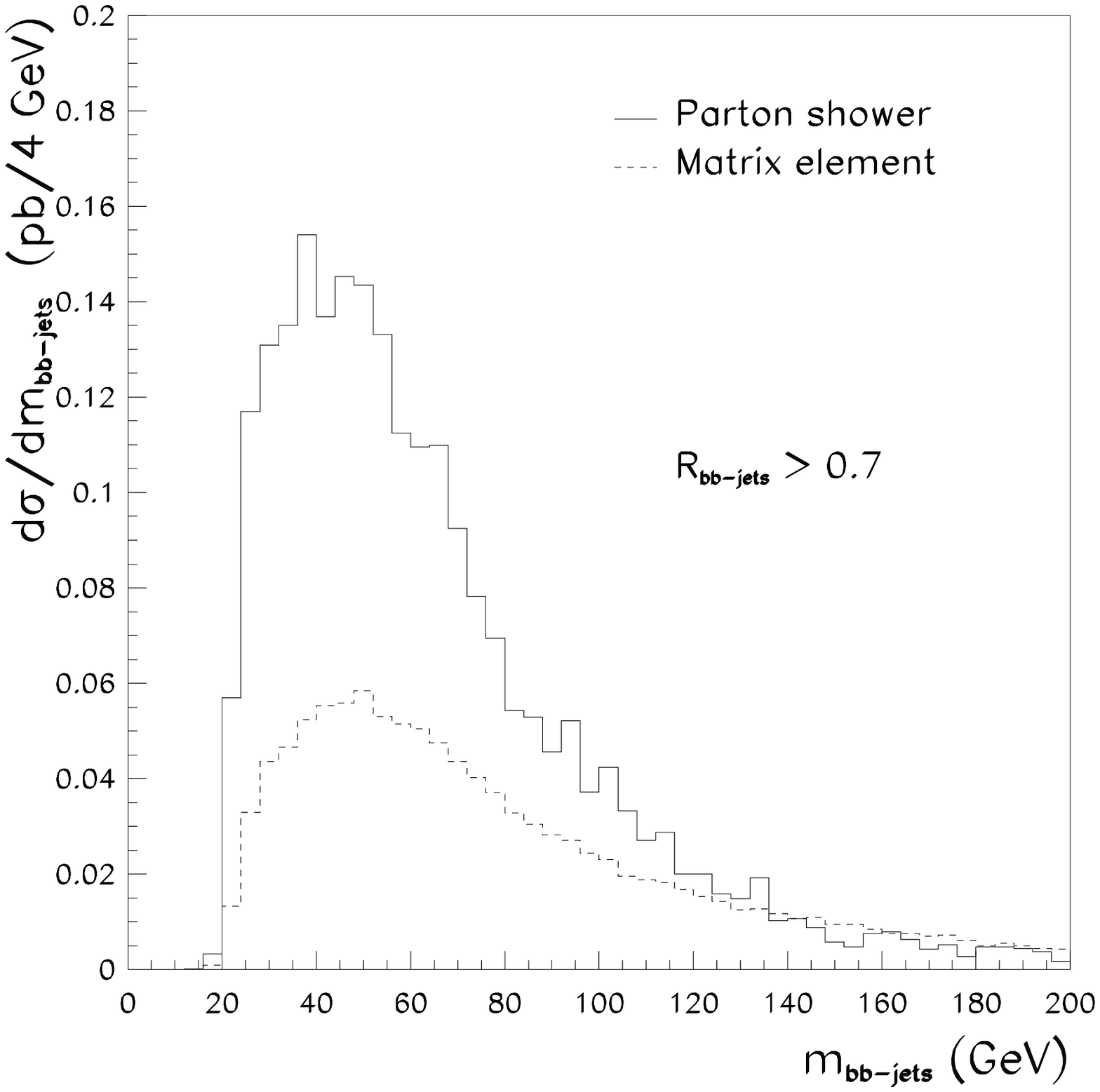,width=4.5cm}\hskip -0.3cm
     \epsfig{file=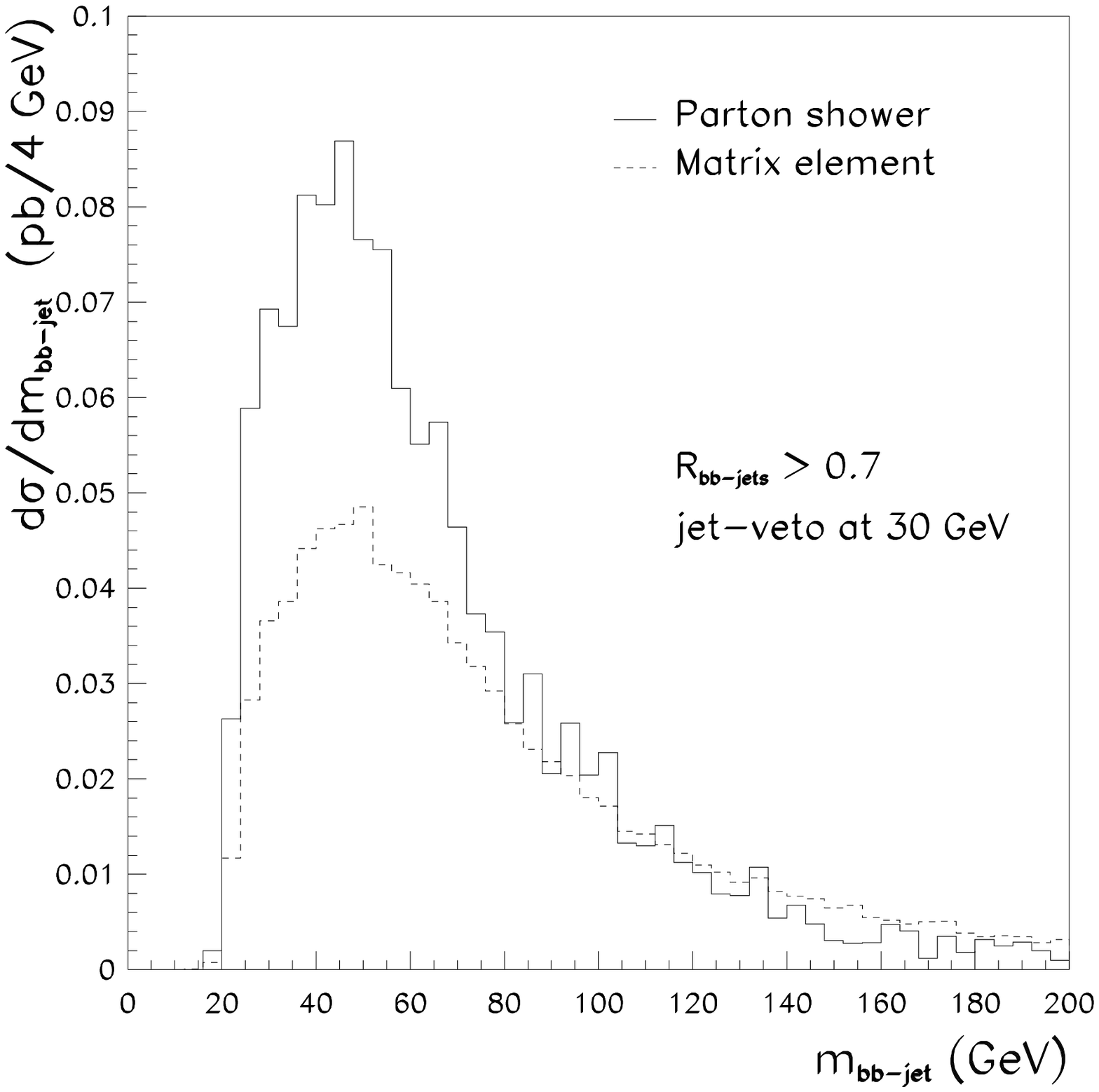,width=4.5cm}

\end{center}
\caption{\em
The invariant mass distribution 
after gradually adding selection requirements is shown.
Solid line denotes the PS events, the dashed one the ME events.
\label{FS2:c}} 
\end{figure}

In Fig.~\ref{FS2:b} the distributions relevant for the experimental analyses are plotted
for events with two reconstructed b-jets and one tagged lepton. One can observe that there
is a significant difference in the predicted cone separation between the b-jets, the
$R_{\rm bb-jets}$\footnote{Cone separation is calculated as the separation in the plane of
pseudorapidity ($\eta$) and azimuthal angle ($\phi$), the $R_{\rm bb-jets} = \sqrt{(\Delta
\phi_{\rm bb-jets})^2 + (\Delta \eta_{\rm bb-jets})^2}$.}.  The parton shower approach
predicts more events with a small cone separation.  These events can be rejected if a
threshold on that separation is required. In fact, they are not contributing to the higher
range of the invariant mass of the b-jet pair. As expected, the shapes of the transverse
momenta of the individual b-jets and of the b-jet system are also harder for the ME
events.  Quite different is the slope of the invariant mass distribution of the b-jet
system; for PS events the distribution falls down more rapidly for higher masses.  In
Fig.~\ref{FS2:c} the invariant mass distribution after gradually adding selection
requirements is drawn.  A better agreement is observed when a large cone separation
between b-jets and the jet-veto on the additional jets are required. This could be
explained by the fact that after these requirements the topology of PS events comes closer
to the ME one. As the ME events exhibit lower average multiplicity for reconstructed jets
than the PS events, the latter are suppressed stronger by the jet veto requirement and thus
give cross-section predictions which are even lower than the ones for ME events.  The invariant
mass distribution for ME events however remains harder than the one for PS events. Please
also note the effect of imposing $R_{\rm bb-jets}~>~0.7$ on the shape of the invariant
mass distribution of the b-jet system, which shifts the distribution maximum to about
 40-60~GeV (compare Fig.~\ref{FS2:b} and~\ref{FS2:c}).

More detailed discussions on the quantitative differences between PS and ME approaches for
the $q \bar q \to W (\to \ell \nu) g (\to b \bar b) $ process can be found in
\cite{ATL-COM-PHYS-2001-032}.

\section{ The $\mathbf{Z b \bar b}$ irreducible background}
 
In this section the estimates for the irreducible $Z b\bar b$ background to the Higgs
boson searches in the $ZH$ production, followed by the $H \to b \bar b$ decay, are
discussed. The evaluation is based on two simulation approaches, ME and PS, as specified
below:

\begin{itemize}
\item
 {\bf ME:} Use the $2 \to 4$ matrix element for $gg, q \bar q \to Z/\gamma^*(\to \ell \ell) b
 \bar b$ process as implemented in {\tt AcerMC}. The $\sigma \times BR$ = 26.2~pb  for
 single leptonic flavour $Z/\gamma^*$ decay and for the invariant mass of the lepton pair
 above 60~GeV.  This matrix element represents lowest order contribution to the $\ell \ell
 b \bar b$ final state. The initial (ISR) and final state radiation (FSR) is simulated
 with parton shower of {\tt PYTHIA}, followed by hadronisation to complete the event
 generation.
\item
 {\bf PS:} Use the $2 \to 1 $ matrix element for $q \bar q \to Z/\gamma^* $ process as
 implemented in {\tt PYTHIA}, followed by the ISR.  The $\sigma \times BR$ = 1640 pb for
 single lepton flavour decay and invariant mass of the lepton pair above 60~GeV. Gluon
 splitting in the ISR partonic cascade is the source of b-quarks in the event. The
 implementation of this process includes ISR/FSR modeled with {\it an improved parton
 shower approach} \cite{Sjostrand1998}, \cite{Sjostrand2000} which integrates some higher-order corrections.
\end{itemize}

\begin{figure}
\begin{center}
     \epsfig{file=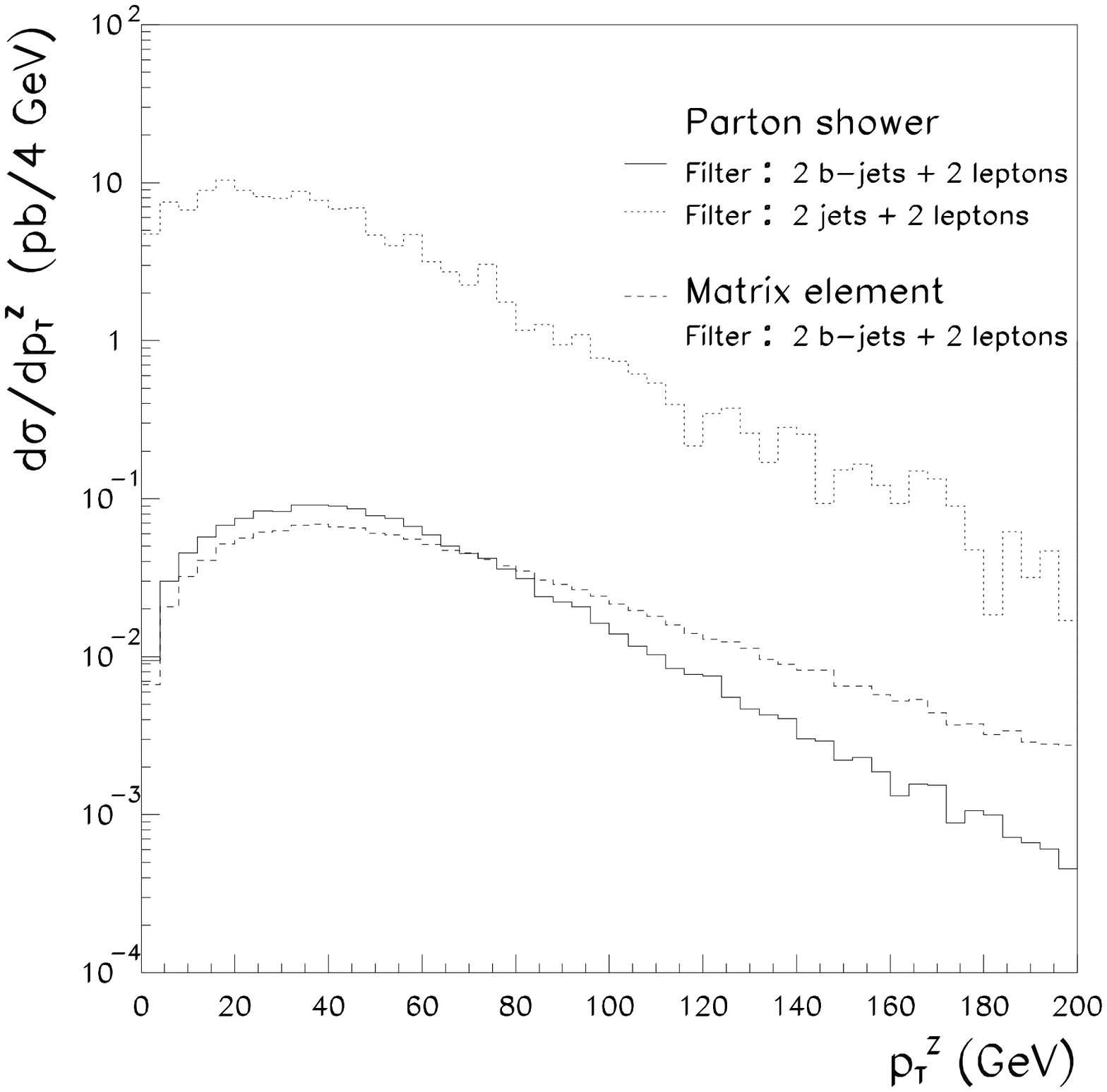,width=4.5cm}\\
     \epsfig{file=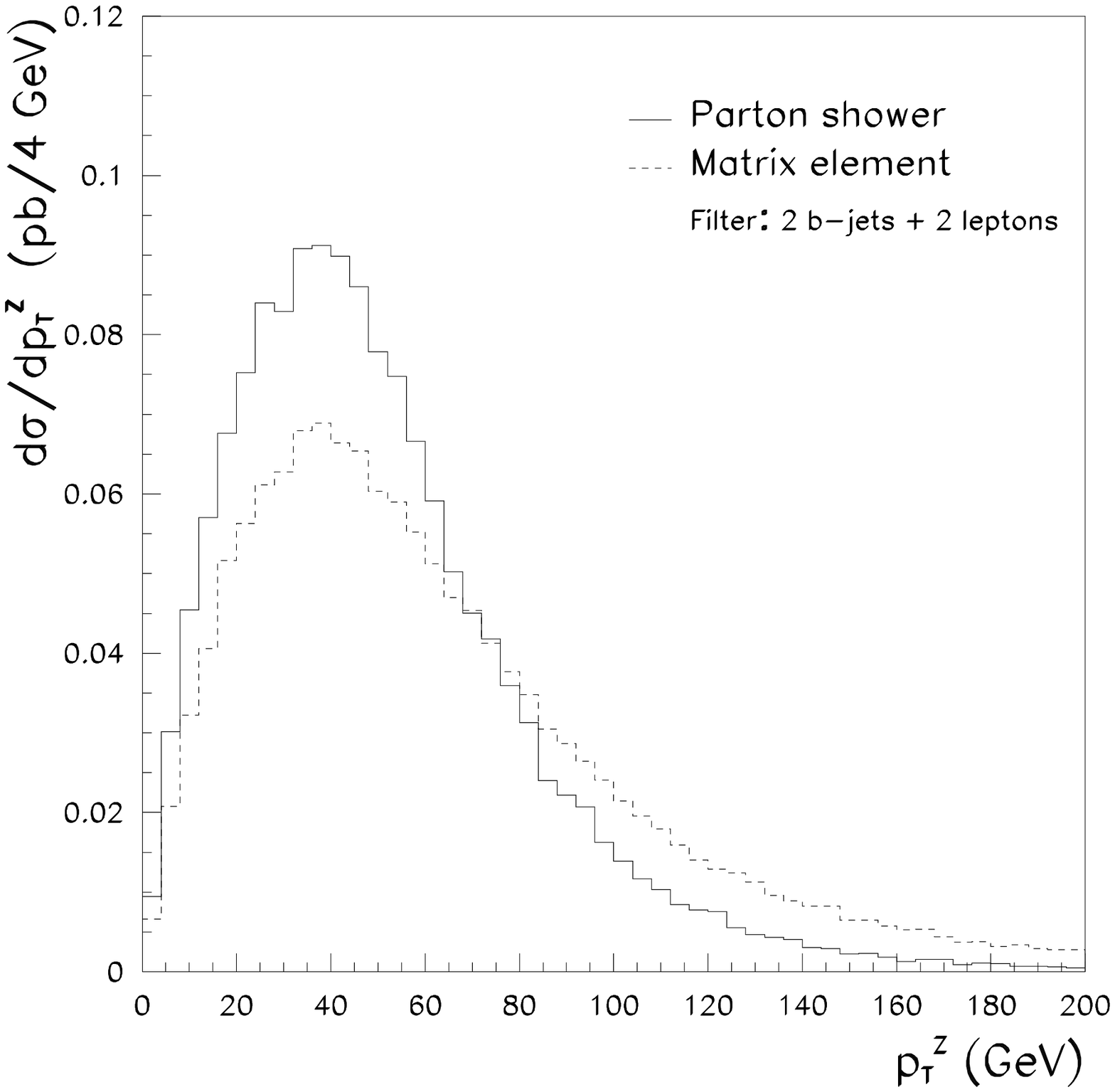,width=4.5cm}\hskip -0.3cm
     \epsfig{file=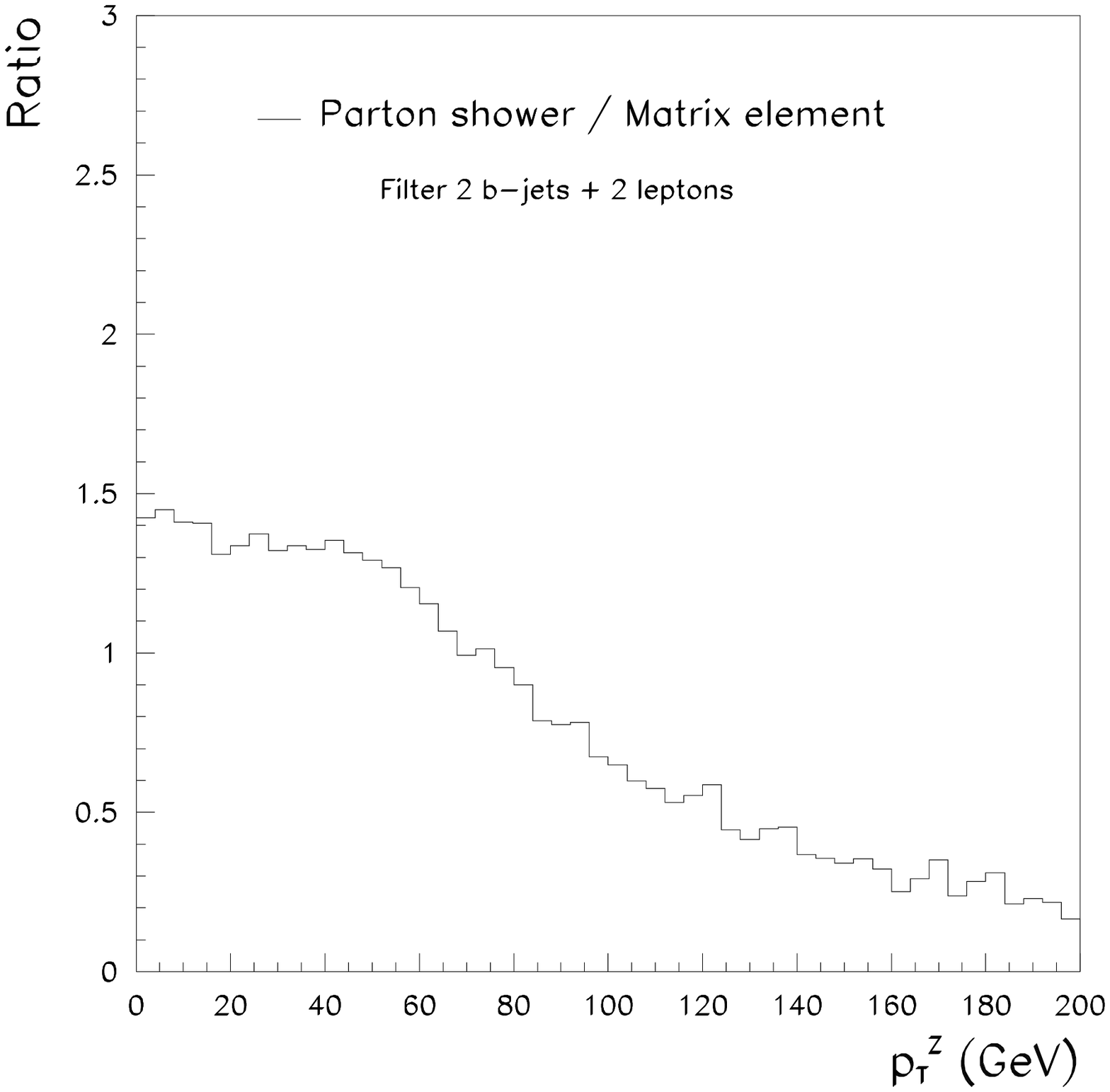,width=4.5cm}
\end{center}
\caption{\em
The transverse momenta of the Z-boson. Solid line denotes the PS events, dashed one
the ME events. Events were filtered as specified in the figure.
\label{FS3:a}} 
\end{figure}

The study begins by comparing differential distributions of the Z-boson transverse momenta
for both simulation approaches, see Fig.~\ref{FS3:a}, after requiring reconstructed
leptons and jets (b-jets) in the final state.  With the generation threshold for the
invariant mass of the lepton pair at 60 GeV, the studied events are dominated by the
on-shell Z-boson exchange.

Contrary to the $W b \bar b$ case, the slopes of the $p_T^Z$ distributions in the ME and
PS events are very different; the ME events are found to be much harder. One should
remember that there are several topologies of the Feynman diagrams leading to the $\ell
\ell b \bar b$ final state, see e.g. \cite{AcerMC}, the dominant one being the contribution
from the multipheral topologies of $gg \to Z/\gamma^* b \bar b$ where each gluon splits
into a $b \bar b$ pair and the $b \bar b$ pair originating from different gluons
annihilates to produce the $Z$-boson. This might explain why in this case the universal
{\it improved parton shower approximation} implemented in {\tt PYTHIA} for $q \bar q \to
Z/\gamma^* $ process is not working as well as in the previous case of the $q \bar q \to W
$ process and the $\ell b \bar b$ events.

\begin{table} 
\newcommand{\lstrut}{{$\strut\atop\strut$}}
  \caption {\em Cross-section for the $gg, q \bar q \to \ell \ell b \bar b$ ME and $q \bar
  q \to Z/\gamma^* $ PS events with $Z/\gamma^* \to \ell \ell$ decay (single
  flavour). Throughout the study the mass of the lepton-pair is required to be above
  60~GeV. Efficiencies for b-tagging and lepton identification are not included, only the
  efficiencies for jet reconstruction and b-jet tagging are taken into account.
\label{TS3:a}} 
\vspace{2mm}  
\begin{center}
\begin{tabular}{lllll}
\hline\noalign{\smallskip}
Selection &  $gg, q \bar q \to \ell \ell b \bar b $ & $q \bar q \to Z/\gamma^*
          (\to \ell \ell) $ \\
          & ME   &  PS \\
\noalign{\smallskip}\hline\noalign{\smallskip}
Generated:  $\sigma \times BR$  & 26.2 pb  & 1640 pb    \\
\noalign{\smallskip}\hline\noalign{\smallskip}
\parbox[c]{3cm}{Two b-jets +\\ two leptons}  &  1.70  pb  & 1.62 pb \\ 
\noalign{\smallskip}
$m_{\ell \ell} = m_Z \pm 10$ GeV  & 1.54  pb  & 1.48 pb  \\
\noalign{\smallskip}
\parbox[c]{3cm}{$m_{\rm bb-jets} = \\100-140$ GeV} & 0.28  pb  &  0.31 pb  \\
\noalign{\smallskip}\hline
\end{tabular} 
\end{center}
\end{table}

\begin{figure}
\begin{center}
     \epsfig{file=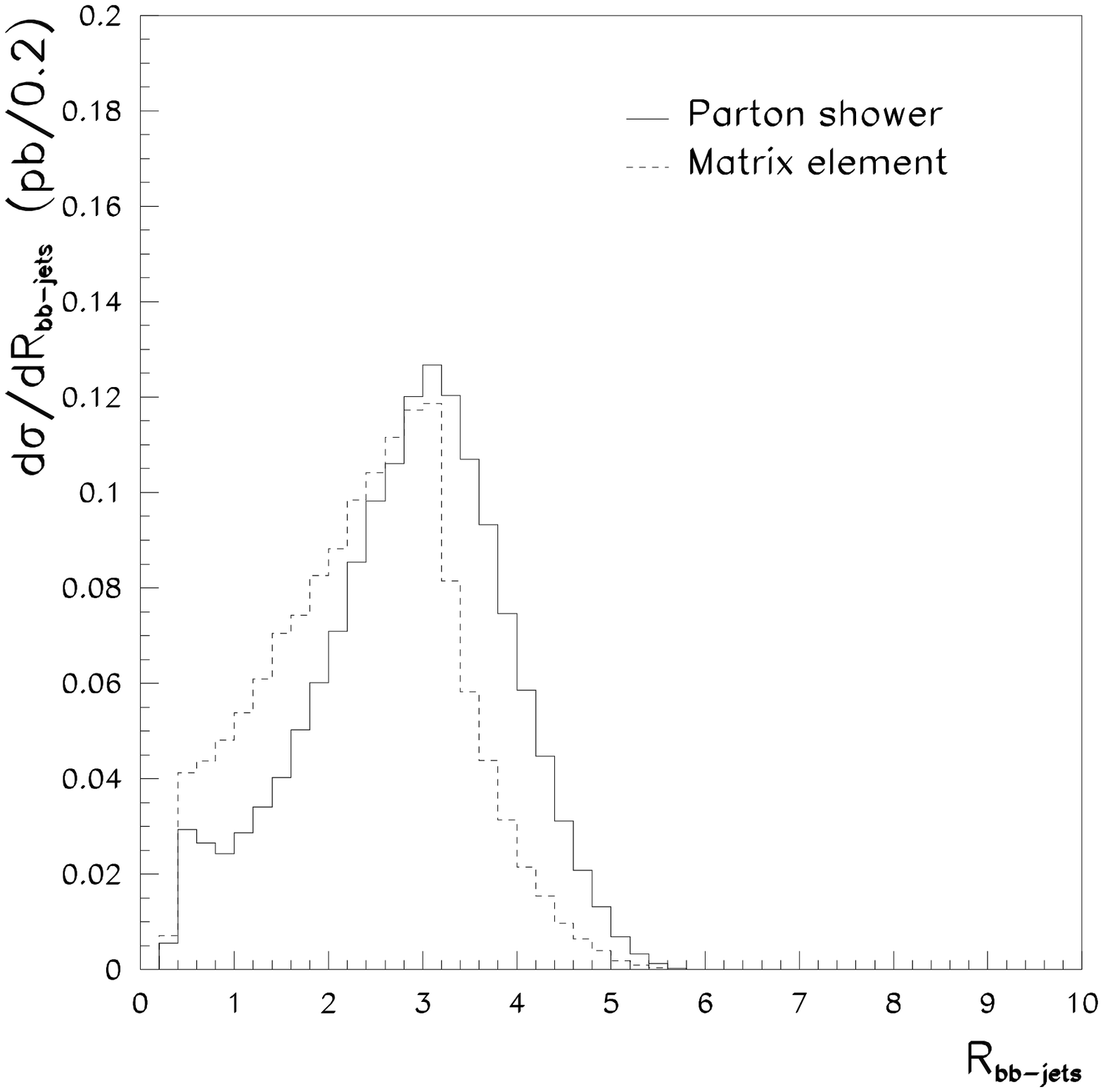,width=4.5cm}\hskip -0.3cm
     \epsfig{file=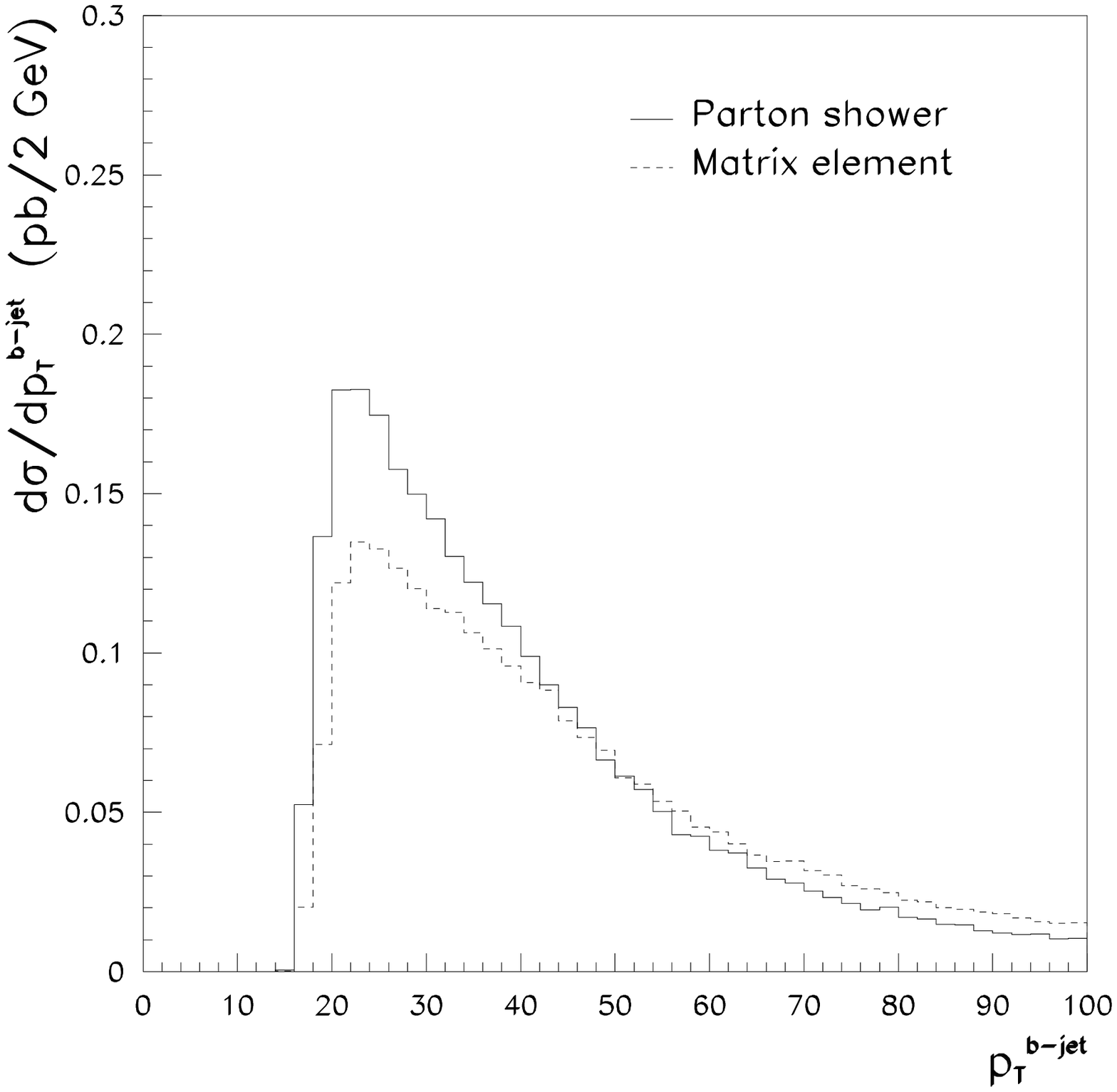,width=4.5cm}\\
     \epsfig{file=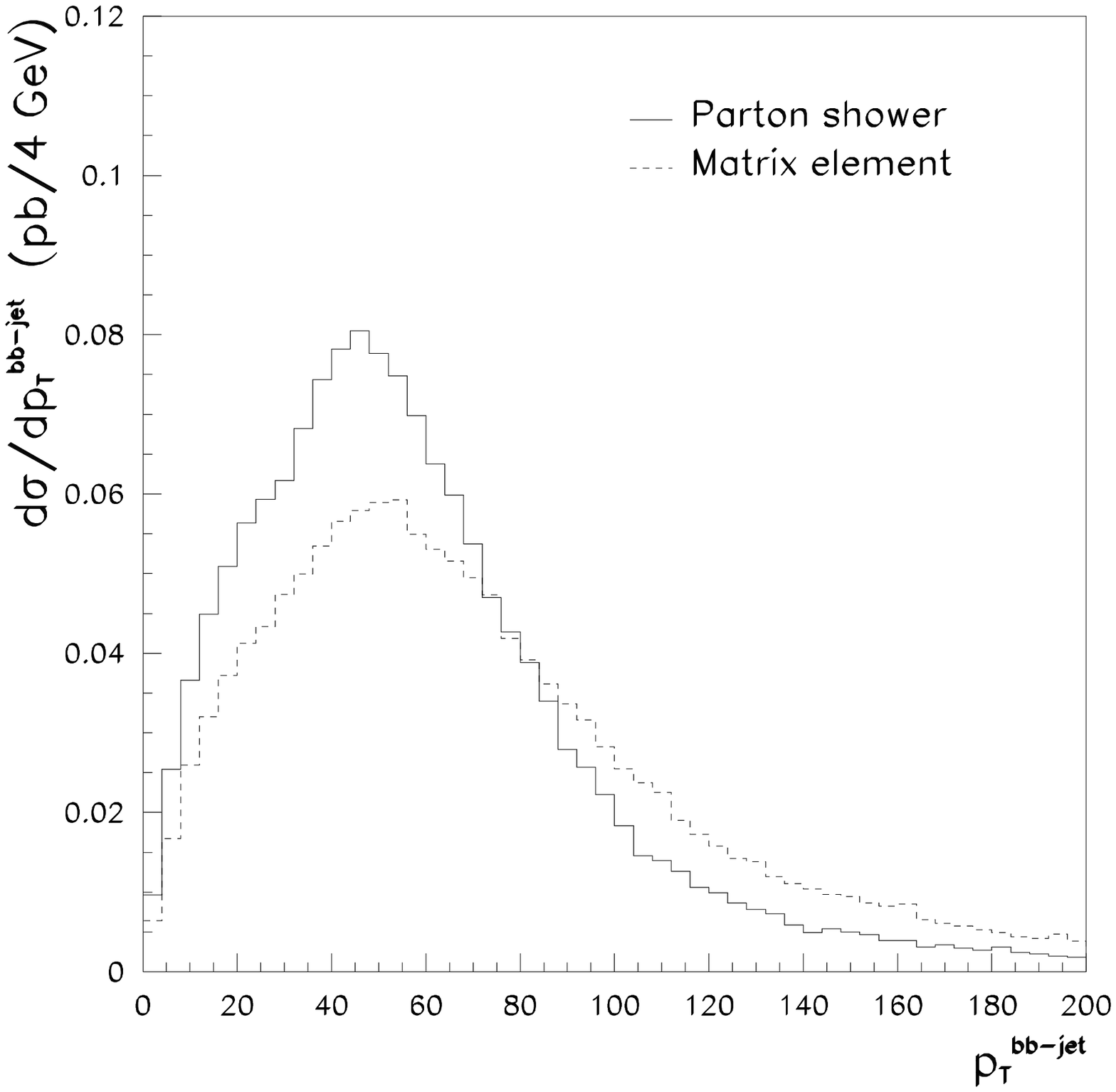,width=4.5cm}
\end{center}
\caption{\em
Distributions of events with two tagged b-jets and a lepton pair with invariant mass
around the mass of the Z-boson: The cone separation between b-jets, transverse momenta of
individual b-jets, and transverse momenta of the b-jet system. Solid line denotes the ME
events, the dashed one the PS events.
\label{FS3:b}} 
\end{figure}

The overall normalisation prediction of both simulation streams, see Table~\ref{TS3:a},
seems to be in reasonable agreement for $\ell \ell b \bar b$ events.  This agreement
remains after requiring lepton-pair and b-jet pair within the mass window, but would
deteriorate if experimental analysis became sensitive to the hard tail of the $p_T^Z$
distribution. In both generation approaches the factorisation/renormalisation energy scale
$Q^2=m_Z^2$ was used.  Also in this case, the variation of the cross-section with the
factorisation/renormalisation scales implemented in {\tt AcerMC} is less than 20\%.

In Fig.~\ref{FS3:b} the distributions relevant for the experimental analyses are
drawn. One can observe that, contrary to the previous case, there is no significant
difference in the cone separation between b-jets ($R_{\rm bb-jets}$) distribution.  In both simulation
approaches the dominant fraction of events has a b-jet pair with a large cone
separation. This marks the topology of $\ell \ell b \bar b$ process as quite different from
$ \ell b \bar b$ events. Furthermore, the shapes of the transverse momenta of the individual
b-jets and of the b-jet system are in this case also quite similar (compatible) for ME and PS events.

\begin{figure}
\begin{center}
     \epsfig{file=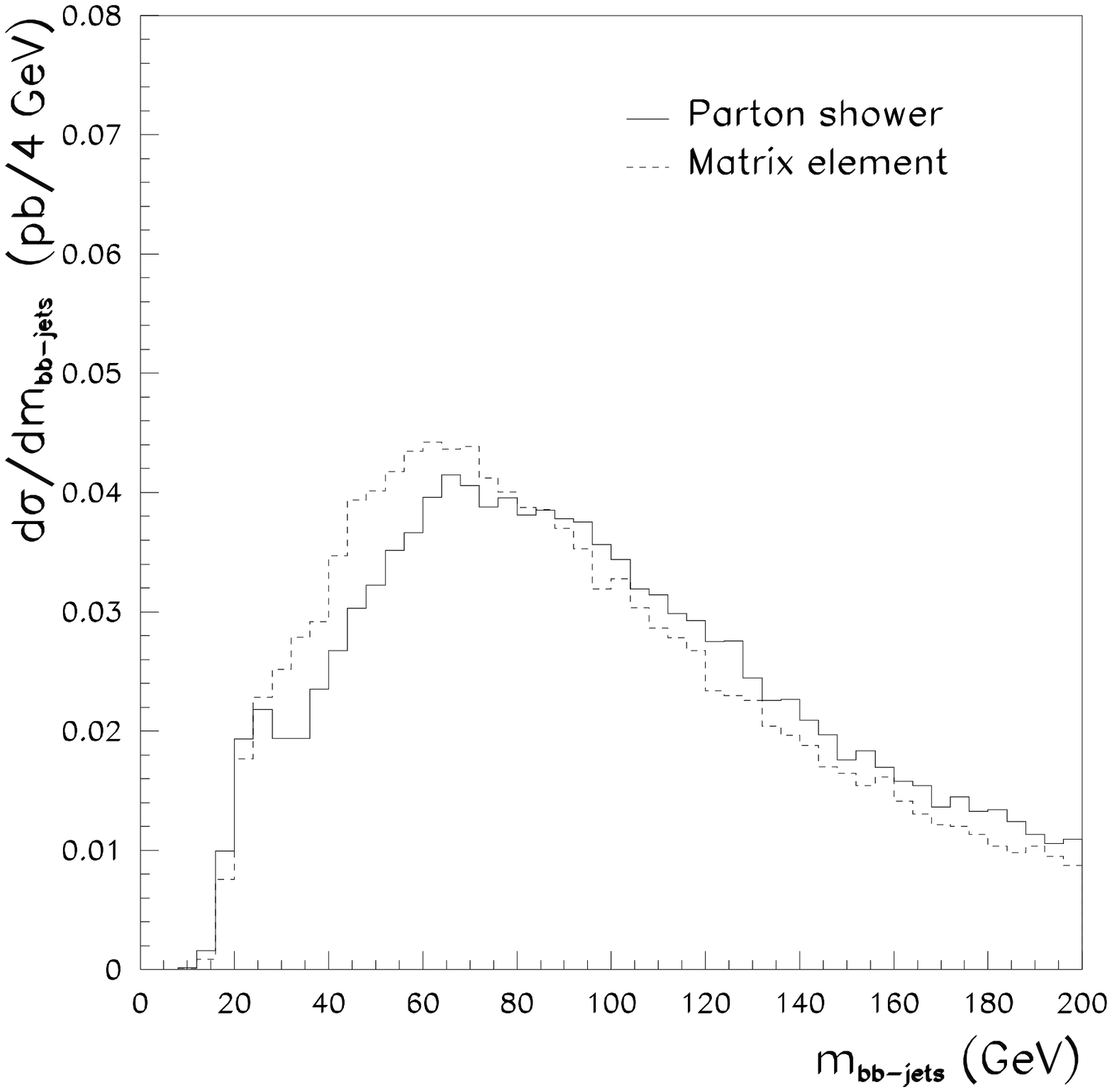,width=4.5cm}\hskip -0.3cm
     \epsfig{file=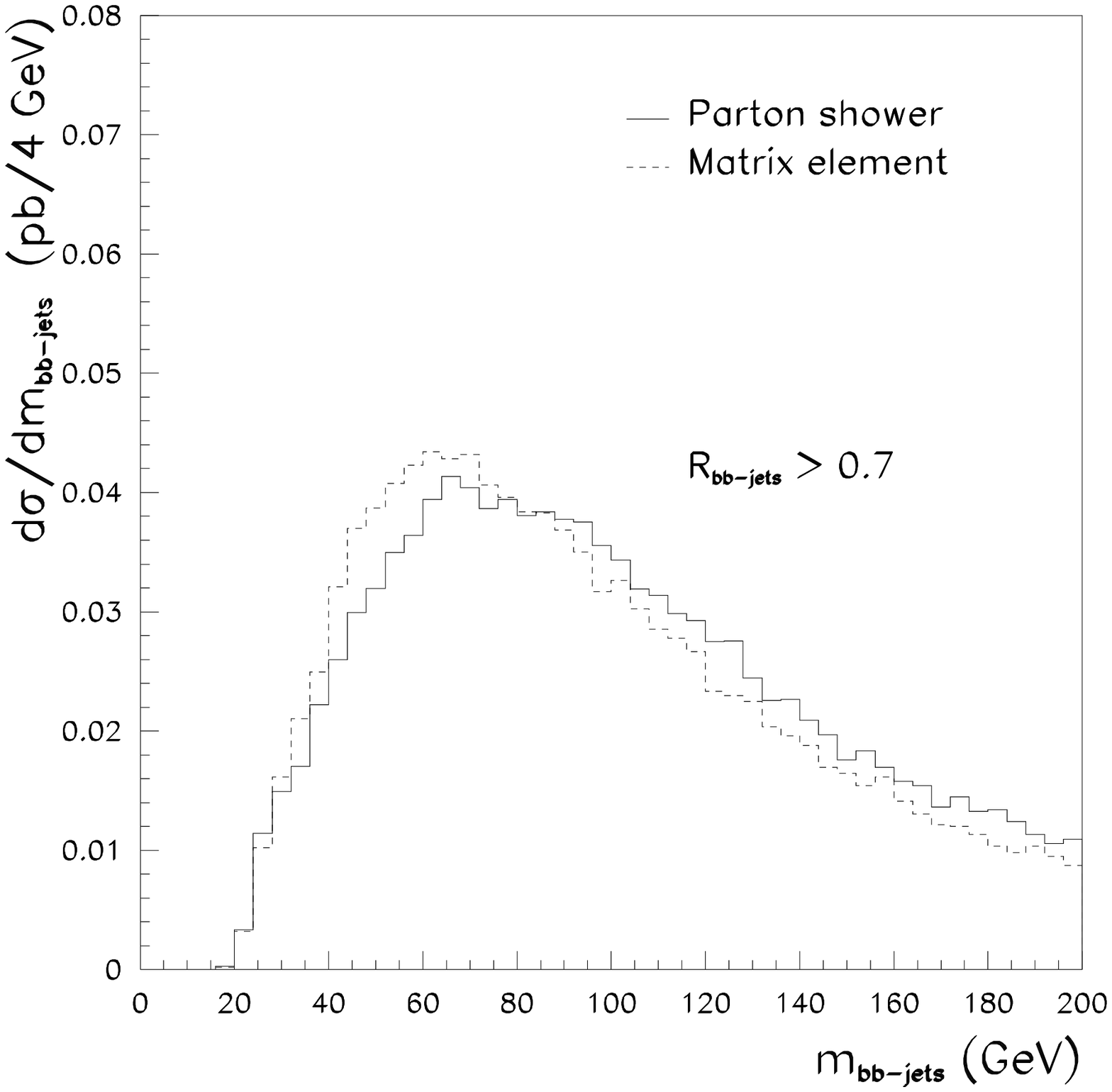,width=4.5cm}
\end{center}
\caption{\em
The distribution of the invariant mass of the b-jet system for events with two b-jets and
lepton pair within the Z-boson mass window is shown.  Solid line denotes the ME events,
the dashed one the PS events.
\label{FS3:d}} 
\end{figure}

\begin{figure}
\begin{center}
     \epsfig{file=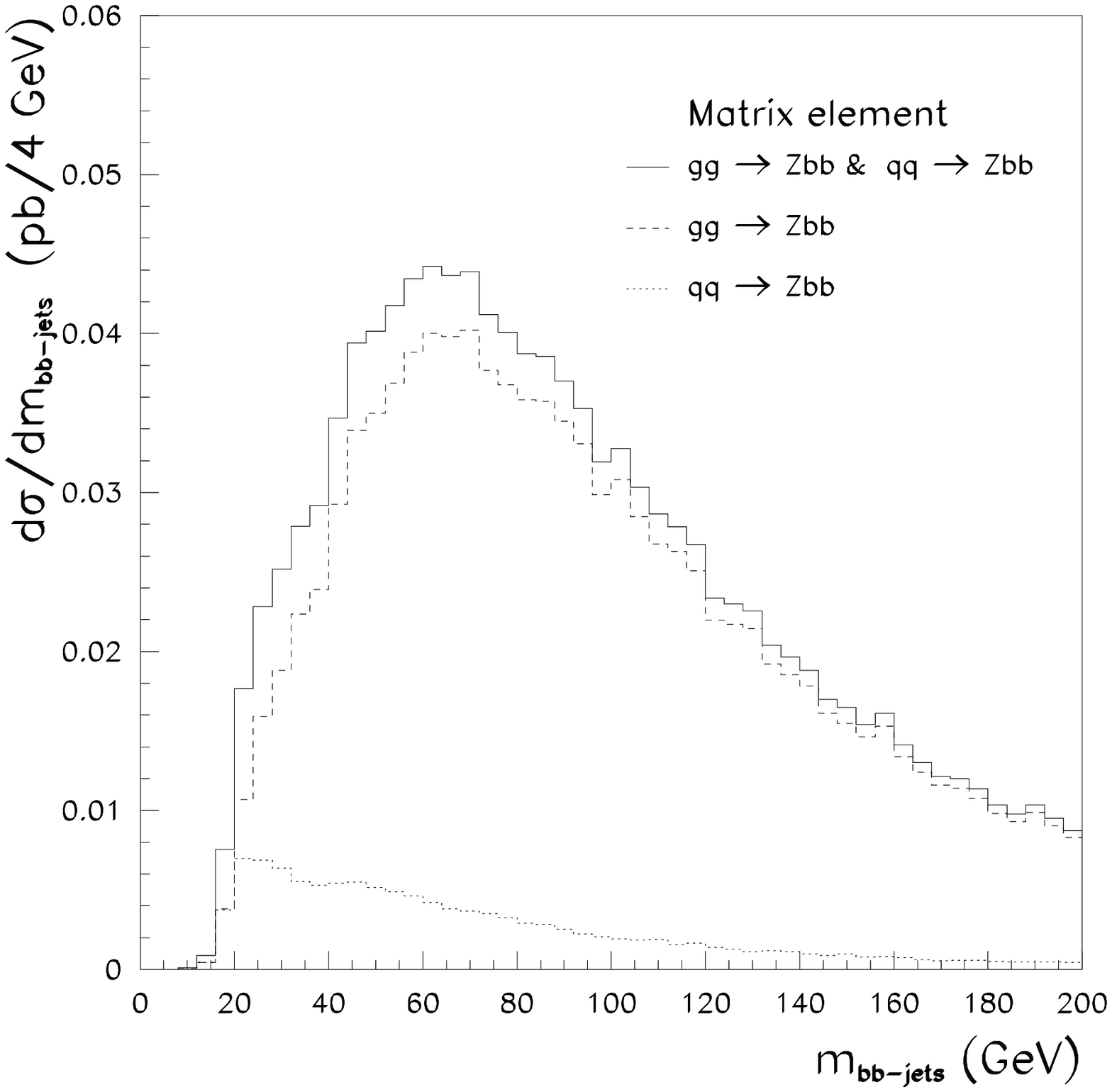,width=4.5cm}\hskip -0.3cm
     \epsfig{file=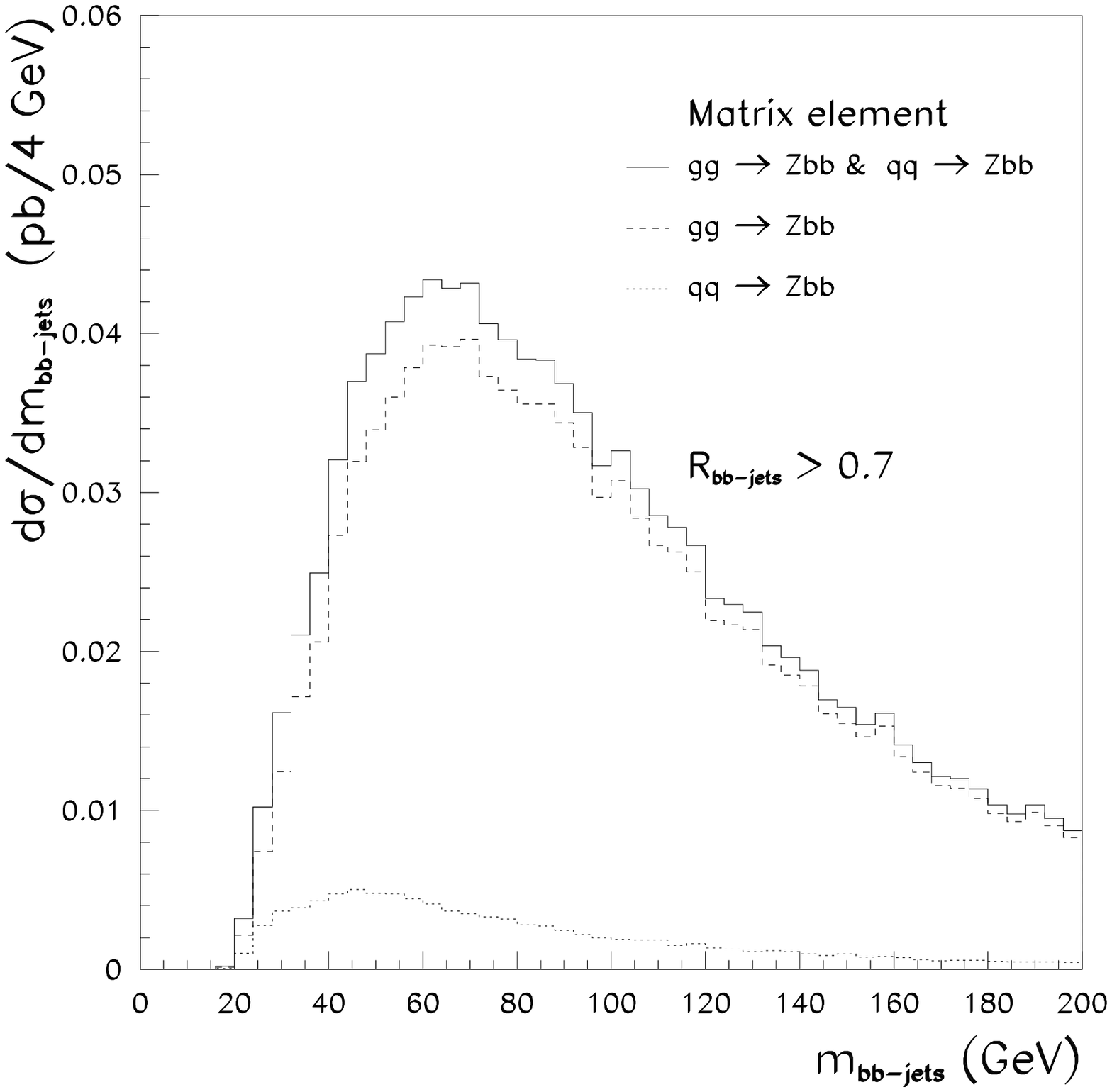,width=4.5cm}
\end{center}
\caption{\em
The distribution of the invariant mass of the b-jet system for events with two b-jets and
lepton pair within the Z-boson mass window, generated with ME, is shown. The $gg$ and
$q\bar q$ components are shown separately.
\label{FS3:e}} 
\end{figure}

Surprisingly similar, given the complexity of the to\-po\-lo\-gies introduced by the Feynman
diagrams, is the distribution of the invariant mass of the b-jet system in PS and ME
events. This is illustrated in Fig.~\ref{FS3:d}. To illuminate this further,
Fig.~\ref{FS3:e} shows the separate ME contributions from the $q \bar q$ and $gg$ events
to the total invariant mass spectrum of the b-jet system.  One sees a distinctly different
shape of both components; as expected the $q \bar q$ component is very similar to the one 
of $W b \bar b$ events.  Nevertheless, the latter class of events contributes only on
the level of 10\% to the total.

\section{ The $\mathbf{t \bar t  b \bar b}$ irreducible background}

In this section the estimates for the irreducible $t \bar t b\bar b$ background
to the Higgs-boson searches in the $t \bar tH$ production, followed by the $H \to b \bar b$
decay, are discussed. Two generation approaches, ME and PS, which lead to the $t \bar t b \bar b$
final state are considered.
 
\begin{itemize}
\item
 {\bf ME:} Use the $2 \to 4$ matrix element for $gg, q \bar q \to t \bar t b \bar b$
 processes as implemented in \cite{AcerMC}. For the QCD component the $\sigma \times BR$ =
 2.7~pb, with leptonic decay (electron or muon) of one W-boson and hadronic decay of the
 second one, both W bosons being produced in the top decays. For the EW component, $gg,
 \to (Z/W/\gamma^* \to) b \bar b$, the $\sigma \times BR$ = 0.26~pb.  These matrix
 elements represent the lowest order contribution to the $t \bar t b \bar b$ final
 state. The interference between QCD and EW component is not available in the present
 implementation of \cite{AcerMC}. In the numerical evaluation the {\it central value} of
 the factorisation scale \cite{NLOttH}, $Q^2_{QCD}~=~(m_t + m_H/2)^2$, with $ m_H = 120$
 GeV was used for the QCD component and the $Q^2~=~m_Z^2$ was used for electroweak one.
 For this process, as already stressed in \cite{AcerMC}, different choices
 of the factorisation scale could lead to the cross-section estimates differing even by
 factor of four.  Event generation is completed by ISR/FSR and hadronisation as modeled
 in {\tt PYTHIA}.
\item
 {\bf PS:} Use the $2 \to 2 $ matrix element for $gg, q \bar q \to t \bar t $ process as
 implemented in {\tt PYTHIA}, followed by the ISR. The $\sigma \times BR$ = 189 pb for leptonic
 decay (electron or muon) of one W-boson and hadronic decay of the second one, both W
 bosons being produced in the top decays.  Gluon splitting in the ISR/FSR partonic cascade
 is the source of additional b-quarks in the event. The default factorisation energy scale of
 {\tt PYTHIA 6.2} is used.
\end{itemize}

There are two classes of processes which lead to the $t \bar t b \bar b$ final state, the QCD and EW
ones; for the corresponding Feynman diagrams see \cite{AcerMC}. The PS events, where the
hard process is just the top-quark pair production, contribute only  the QCD
component. Consequently, only the QCD ME component should be directly compared with the PS
one.

\begin{figure}
\begin{center}
     \epsfig{file=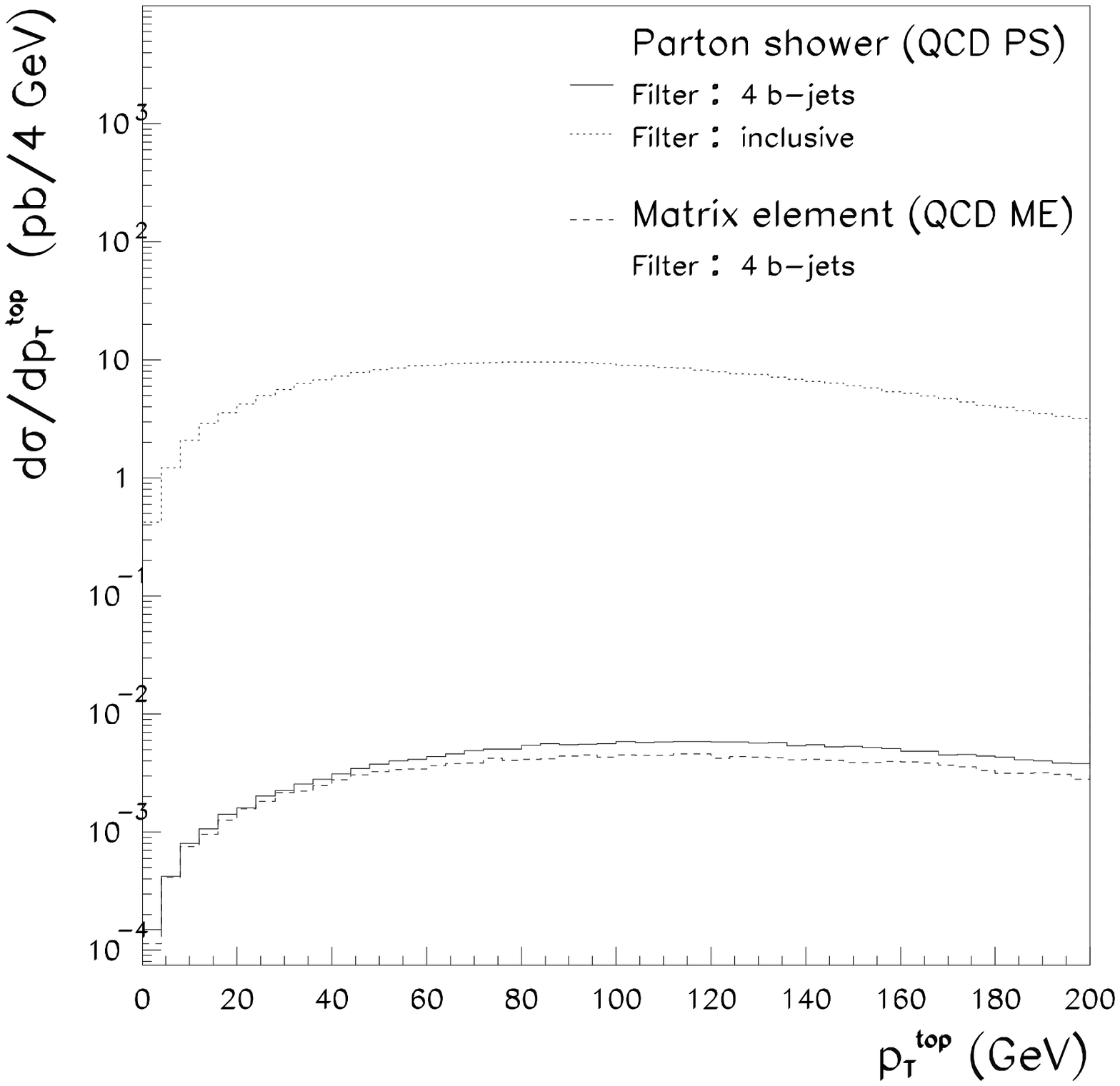,width=4.5cm}\\
     \epsfig{file=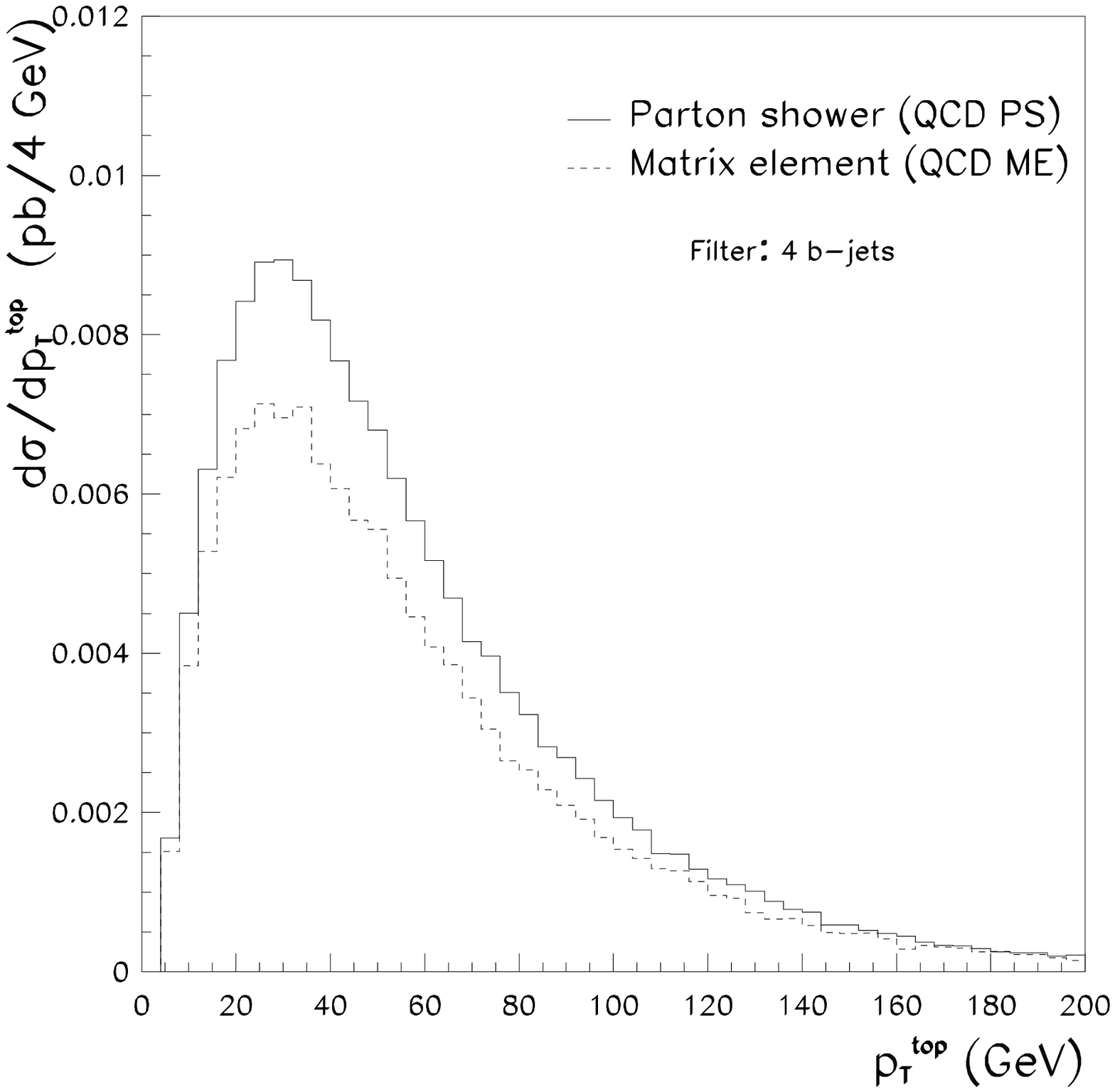,width=4.5cm}\hskip -0.3cm
     \epsfig{file=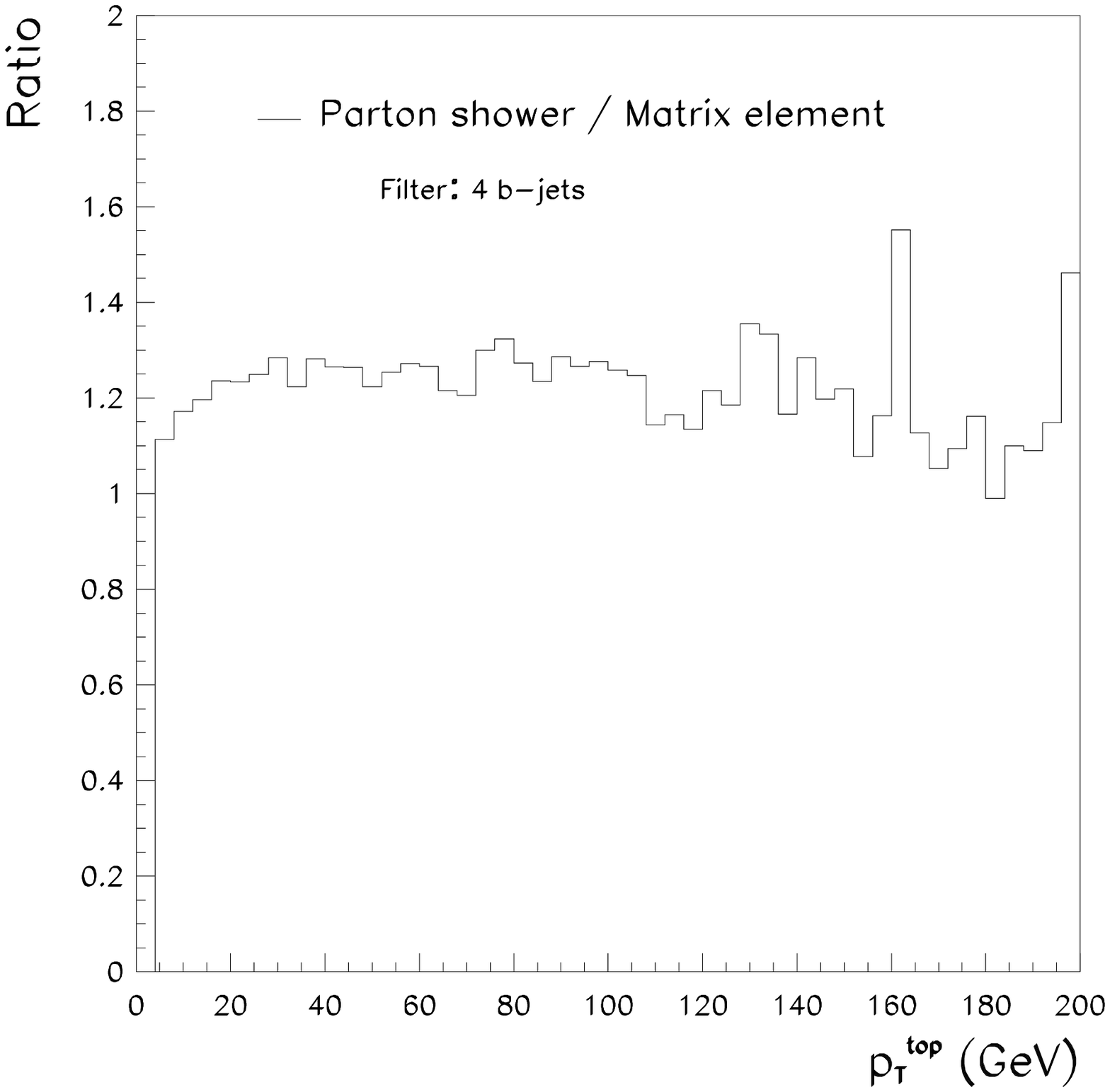,width=4.5cm}
\end{center}
\caption{\em
The  transverse momenta of the top-quarks. Solid line denotes QCD PS events,
dashed line the QCD ME events. Events were filtered as indicated on the plots.
\label{FS4:a}} 
\end{figure}

As an inclusive control distribution the transverse momenta spectra of the top quarks were
chosen. Fig.~\ref{FS4:a} shows that there is quite a good agreement between PS and ME
predictions. With the  factorisation energy scale used for evaluating the ME
predictions the absolute normalisation agrees within 20\% and the ratio of the PS and ME
distributions is amazingly flat. For other choices of the factorisation energy scale, the ratio
would be quite different (see Table with the total cross-sections in \cite{AcerMC}), but
subsequent checks confirm that the distributions remain very similar.

In the proposed experimental analysis \cite{ATL-PHYS-TDR} both top-quarks have to be
reconstructed and the remaining two b-jets are than considered as possible candidates for the
Higgs boson decay products. In the present study this selection procedure was replaced by
the tight matching requirements of the b-jets and their partonic origin.  These lead to a
very clean separation between b-jets originating from the top-quarks and those which are
not. In what follows only the distributions of the b-jets which are not identified as
originating from the top-quark decays are considered.

In Table~\ref{TS4:a} the expected total cross-sections and the cross-sections after simplified
event selection are given. The EW ME component is on the level of 10\% of the total ME
cross-section and on the level of 25\% of all ME events accepted in the mass window.  The fact
that the resonant electroweak background is not negligible makes the prospects for the
observability more difficult, especially for the Higgs boson masses closer to the mass of
the Z-boson. This mass range has been already excluded for the SM Higgs boson \cite{LEP1},
but for the MSSM scenarios is still in the region of a possible discovery \cite{LEP2}.
While resonant EW component was estimated some time ago as negligible in 
\cite{ActaB30}\footnote{It was the consequence 
of the implementation for the $Q \bar Q Z$ process in {\tt PYTHIA 5.7}
generator used at that time, which was not evaluating the total cross-section correctly.},
the non-resonant EW component was to our knowledge not considered prior to this study. It
is clear that the evaluation presented in \cite{ActaB30} should be now revised to include
the EW background properly.

\begin{table} 
\newcommand{\lstrut}{{$\strut\atop\strut$}}
  \caption {\em Cross-sections for the QCD $gg, q \bar q \to t \bar t b \bar b$, and EW
  $gg \to t \bar t b \bar b$ and QCD $gg, q \bar q \to t \bar t $ production PS with one
  W-boson from top-quark decaying leptonically (electron or muon), and the other one
  hadronically.
\label{TS4:a}} 
\vspace{2mm}   
\begin{center}
\begin{tabular}{llll} 
\hline\noalign{\smallskip}
Selection  &  $gg, q \bar q \to t \bar t b \bar b $ 
           &  $gg \to t \bar t b \bar b $
           &  $gg, q \bar q \to t \bar t$ \\
           & (QCD ME)  & (EW ME)   &  ( QCD PS) \\
\noalign{\smallskip}\hline\noalign{\smallskip}
\parbox[c]{2cm}{Generated:\\  $\sigma \times BR$} & 2.7 pb  & 0.26 pb & 189 pb    \\
\noalign{\smallskip}\hline\noalign{\smallskip}
\parbox[c]{2cm}{4 b-jets + \\1 lepton +\\ 2 jets}   & 0.123   pb & 0.014 pb & 0.145 pb \\
\noalign{\medskip}
\parbox[c]{2cm}{$m_{\rm bb-jets} = \\  100-140$ GeV} & 0.013  pb & 0.003  pb  &  0.014 pb  \\
\noalign{\smallskip}\hline
\end{tabular} 
\end{center}
\end{table}

\begin{figure}
\begin{center}
     \epsfig{file=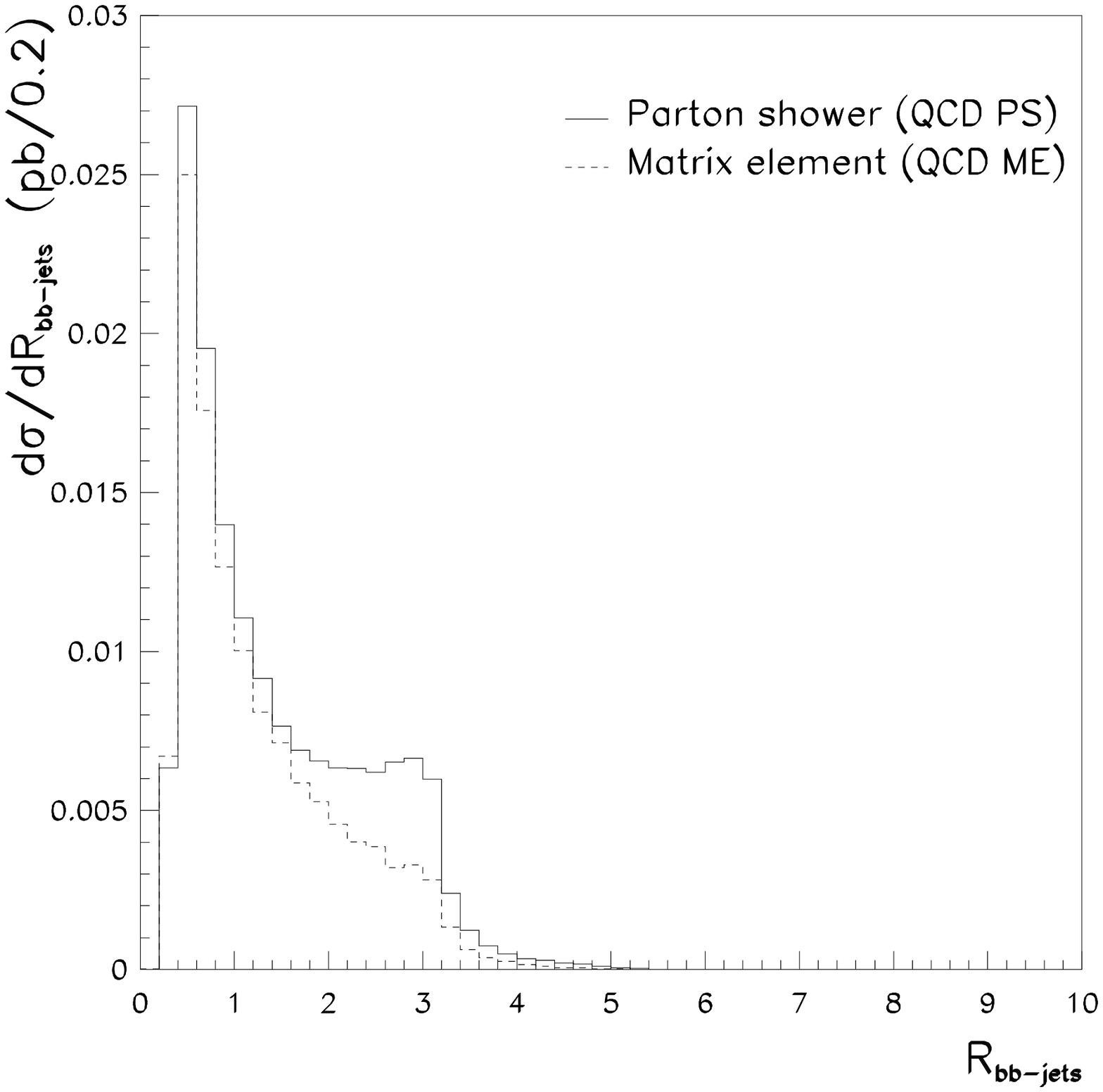,width=4.5cm}\hskip -0.3cm
     \epsfig{file=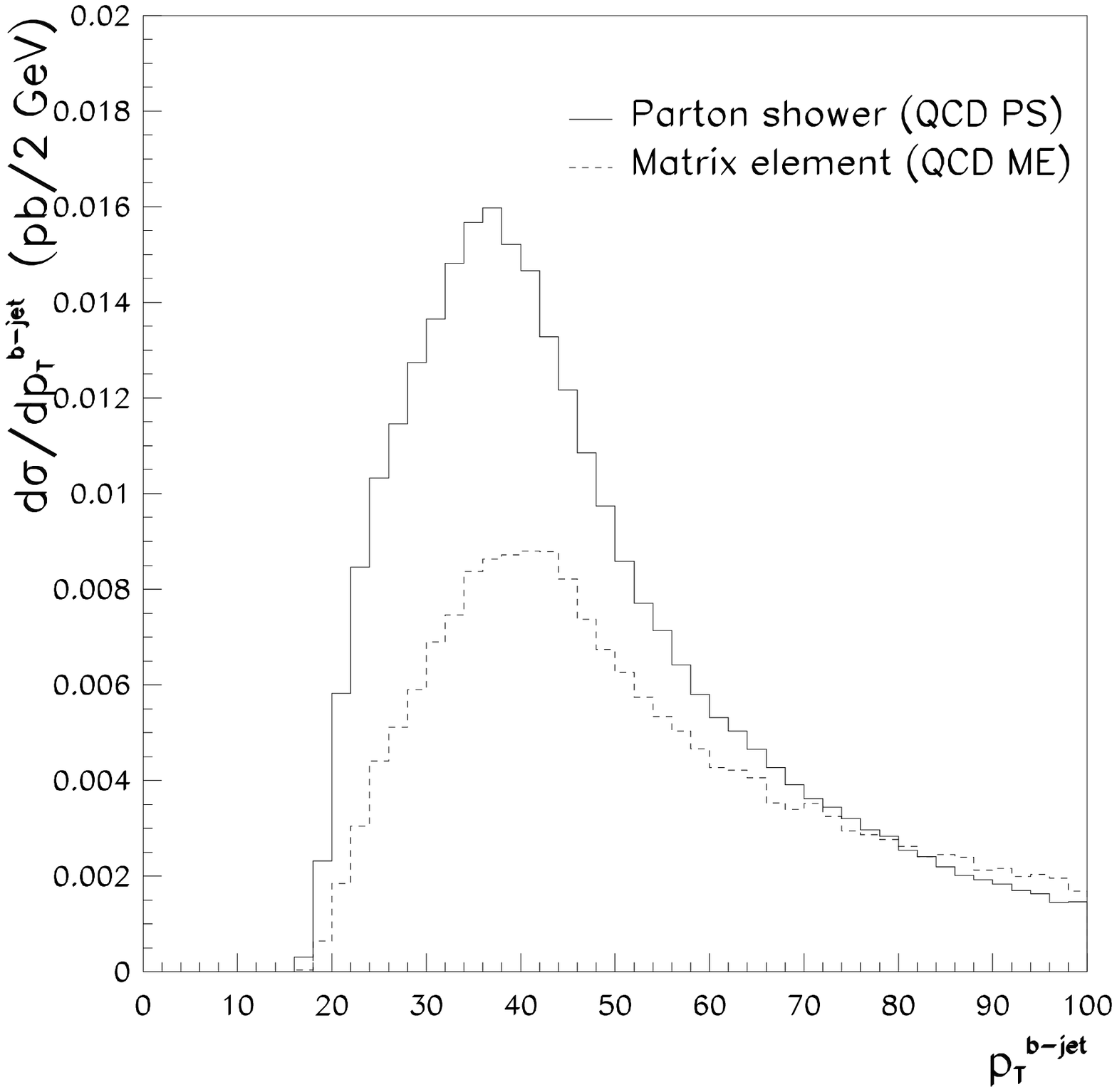,width=4.5cm}\\
     \epsfig{file=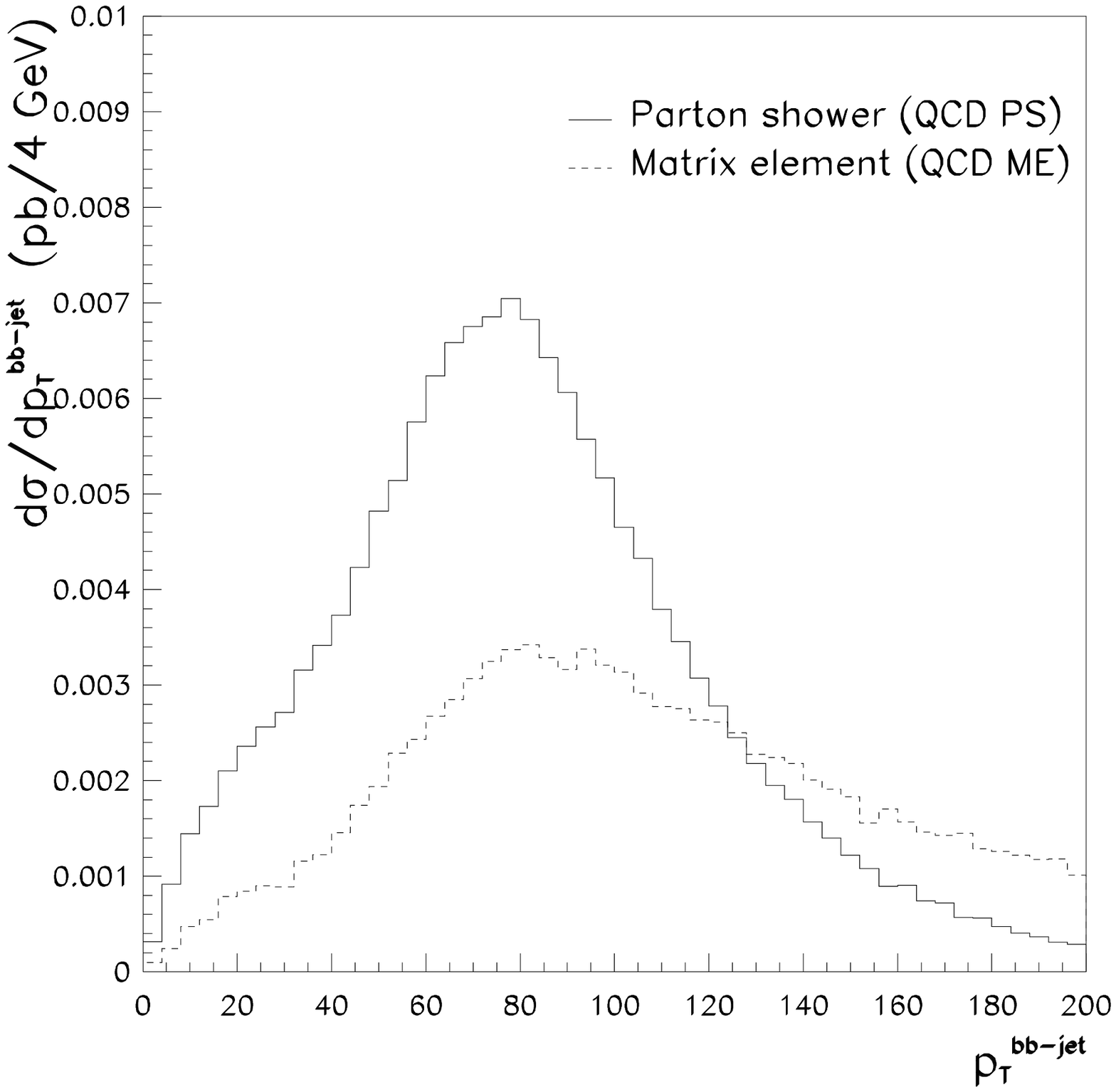,width=4.5cm}
\end{center}
\caption{\em
For events with four b-jets, isolated lepton and at least two light jets, the
distributions for b-jets not originating from top-quark decays are drawn.  Top: The cone
separation between b-jets, transverse momenta of individual b-jets; Bottom: transverse
momenta of the b-jet system.  Solid line denotes the events generated with QCD ME for $t
\bar t b \bar b$ production, the dashed one the ones with QCD PS.
\label{FS4:b}} 
\end{figure}

\begin{figure}
\begin{center}
     \epsfig{file=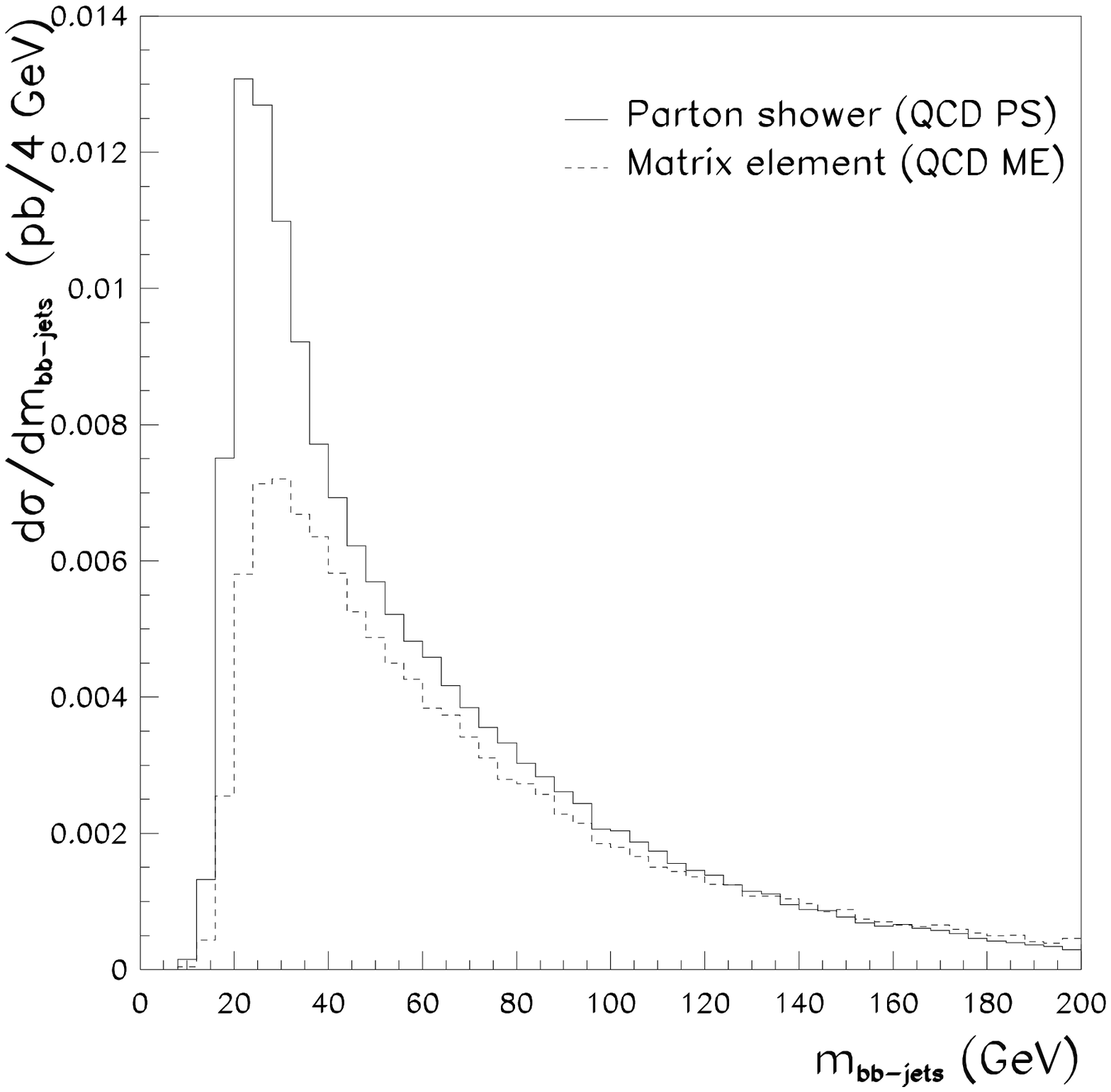,width=4.5cm}\hskip -0.3cm
     \epsfig{file=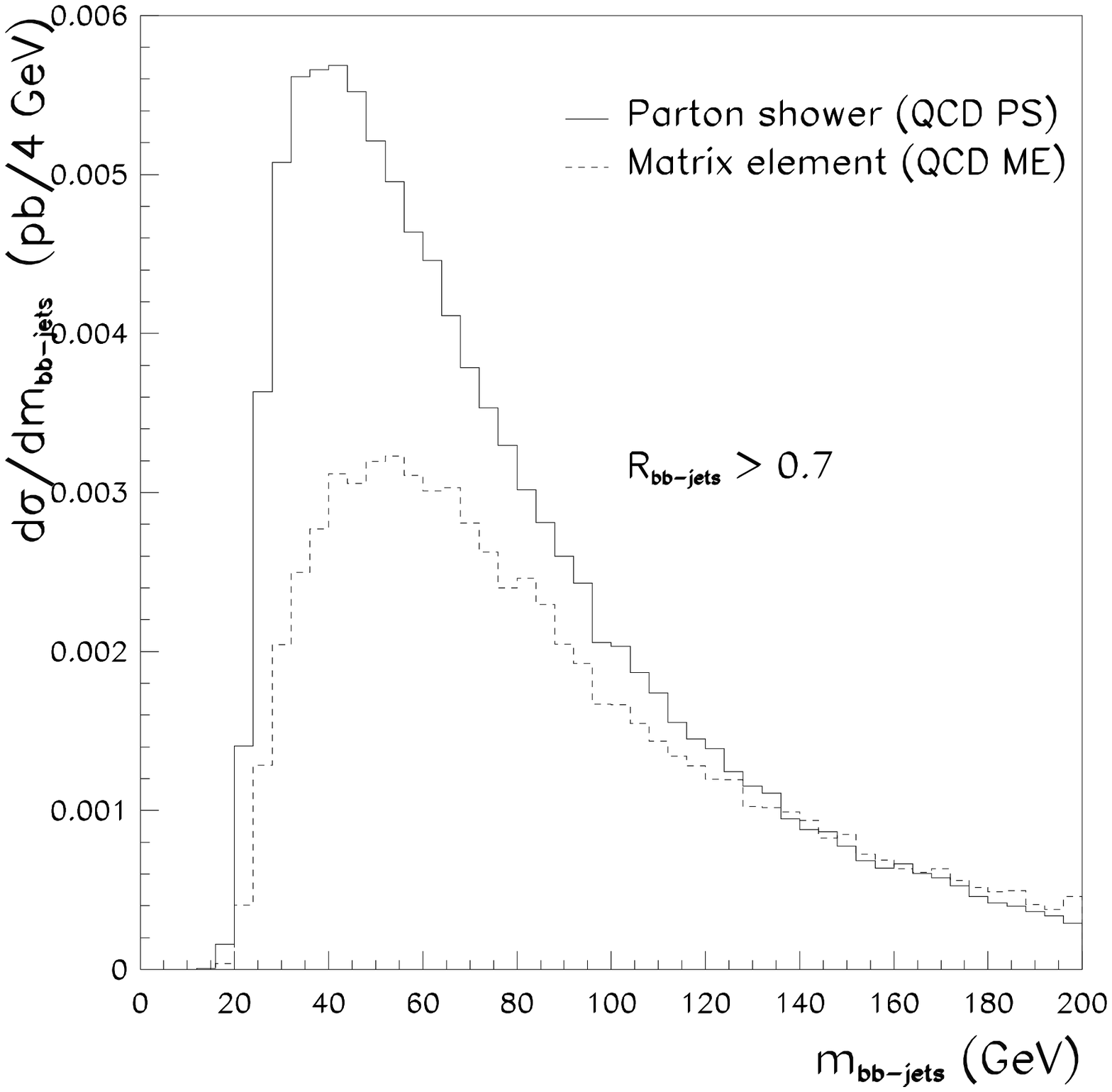,width=4.5cm}
\end{center}
\caption{\em
For events with four b-jets, isolated lepton and at least two light
jets the invariant mass of the b-jet system for b-jets not originating from
top-quark decays is drawn.Solid line denotes the events generated with QCD ME for  
$t \bar t b \bar b$ production, the dashed one the events generated by QCD PS.
\label{FS4:g}} 
\end{figure}

In Fig.~\ref{FS4:b} the distributions relevant for the experimental analyses, simulated
with QCD ME and PS simulation approaches, are drawn. One can observe some differences within
the expected cone separation between b-jets, $R_{\rm bb-jets}$. In both simulation approaches the dominant
fraction of events has the b-jet pair with a small cone separation. Quite different are
the shapes of the transverse momenta of individual b-jets and of the b-jet system. In
particular, the distribution of the transverse momenta of the b-jet system is much harder
in the ME events than in the PS events. Nicely enough, the shape of the invariant mass
distribution of the b-jet system is quite similar in the relevant mass range, see
Fig.~\ref{FS4:g}.  Moreover, the normalisations are in a satisfactory agreement in the
relevant mass range, for lower masses the PS predictions are exceeding the ME ones.

So far only the comparison between ME and PS predictions for the QCD $t \bar t b \bar b$
events was discussed. The EW component can at present only be simulated with the ME
implemented in {\tt AcerMC}. The implementation gives a possibility to estimate either
only the resonant part, the $gg, q \bar q \to t \bar t Z/\gamma^*$ production with
$Z/\gamma^* \to b \bar b$ decay, or to explore the full EW contribution, namely the
process $gg \to t \bar t \to$ $(Z/W/\gamma^* \to) b \bar b t \bar t$. An implementation of
the $q \bar q \to t \bar t$ $\to (Z/W/\gamma^* \to) b \bar b t \bar t$ is still missing,
but will very likely contribute no more than 10-20\% of the total EW background (assuming
the same ratio as for QCD $q \bar q $ and $gg$ contributions).

\begin{figure}
\begin{center}
     \epsfig{file=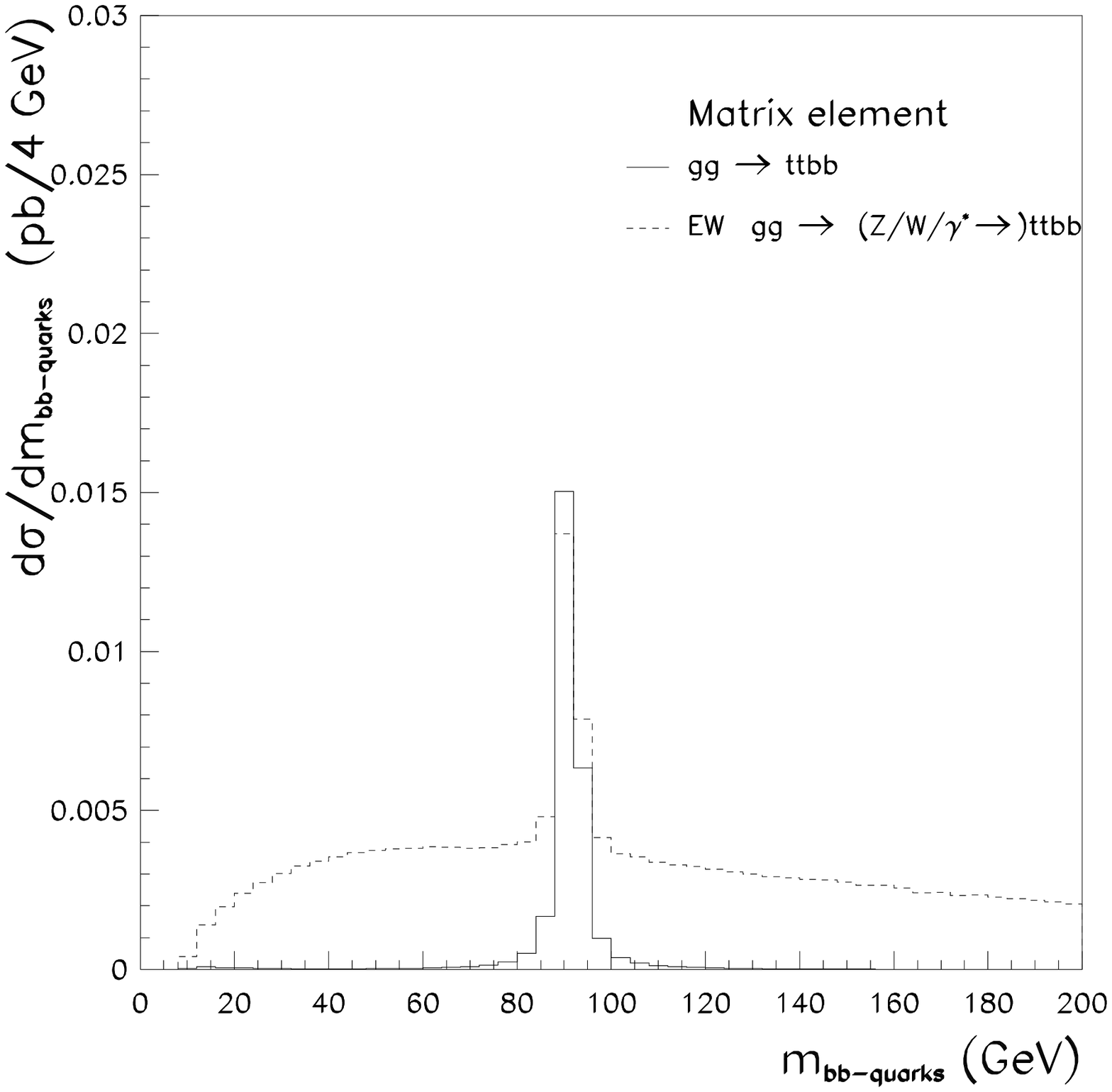,width=4.5cm}\\
     \epsfig{file=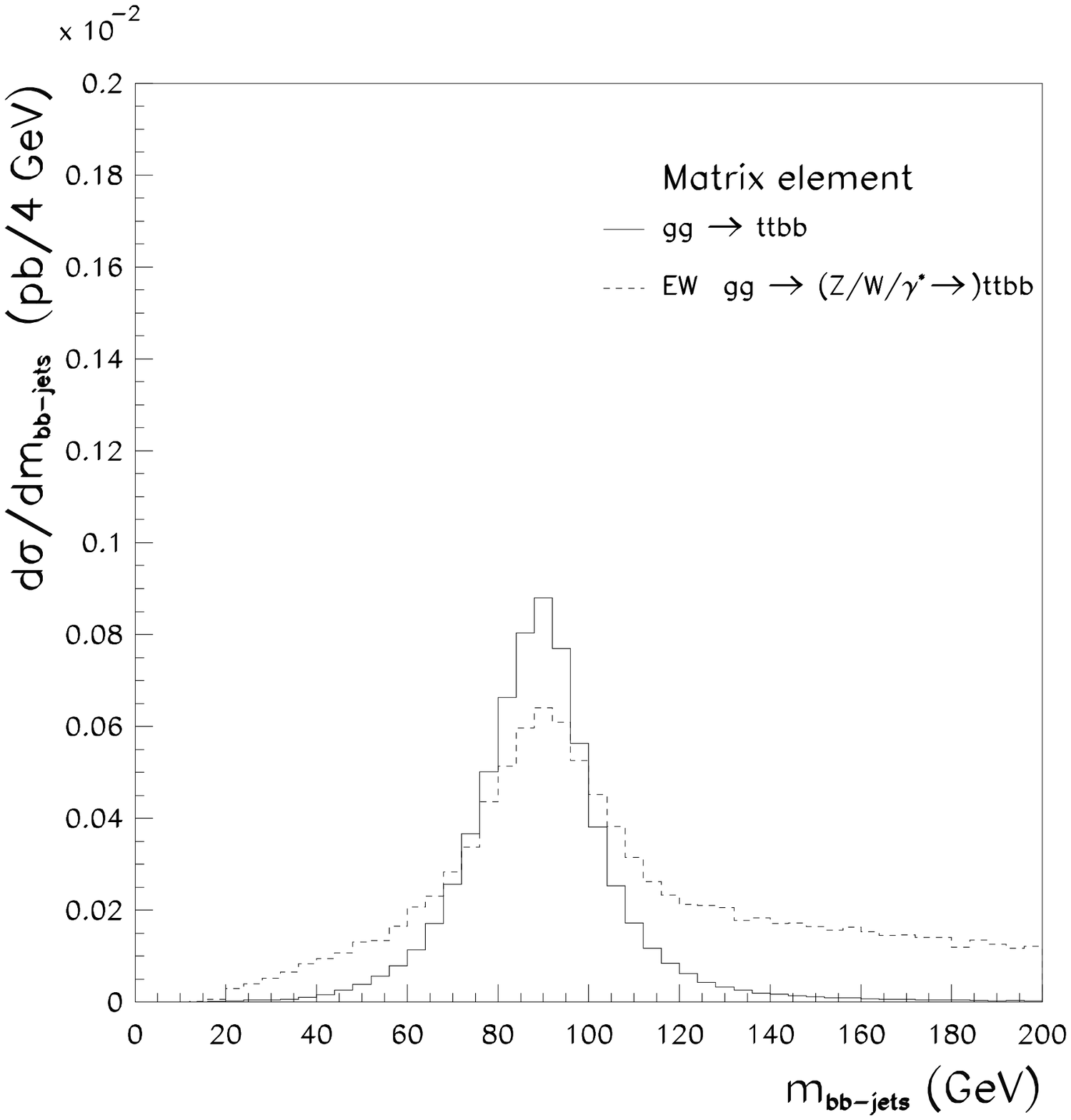,width=4.5cm}\hskip -0.3cm
     \epsfig{file=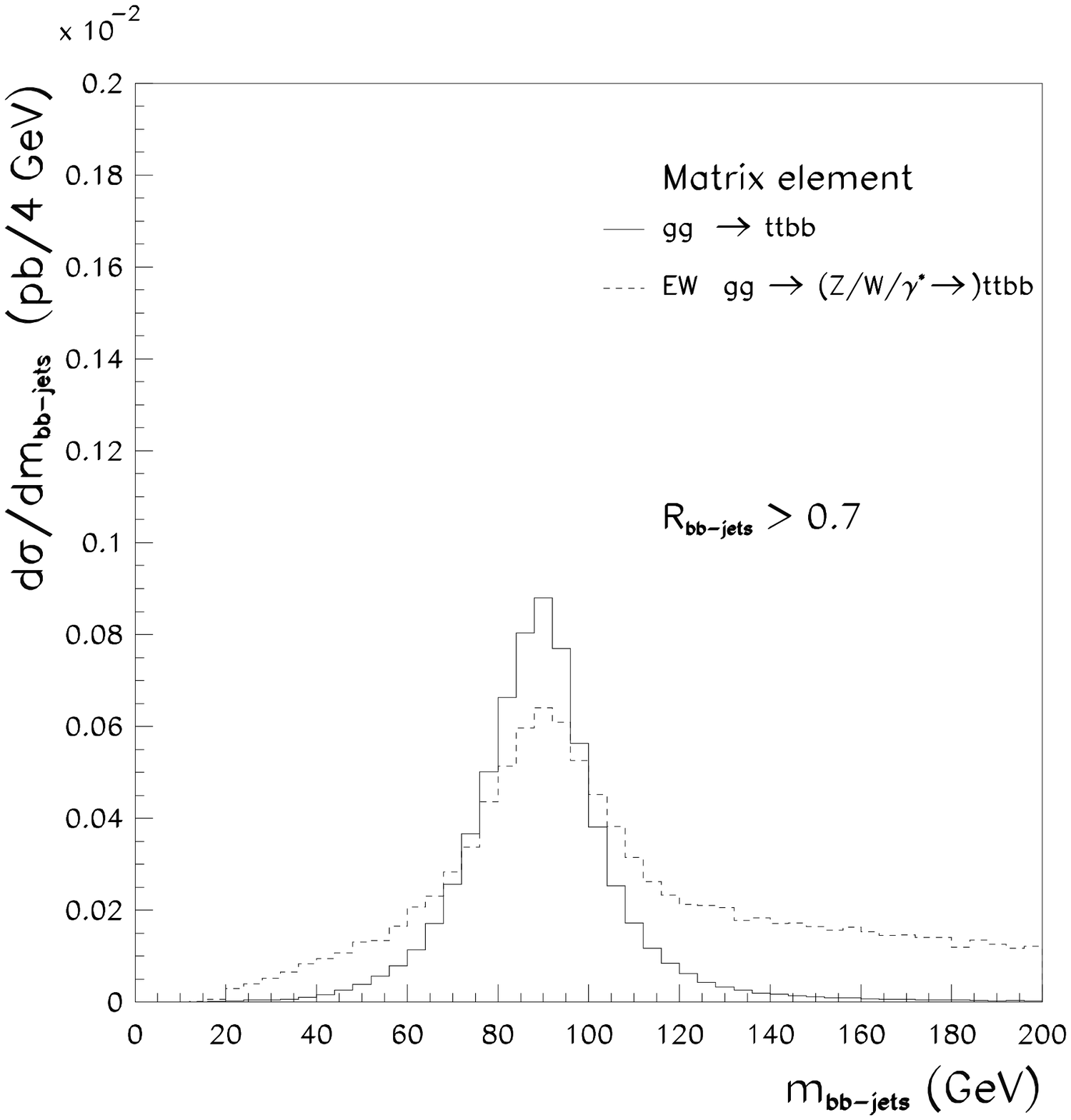,width=4.5cm}
\end{center}
\caption{\em
Invariant mass of the b-quarks system (top plot) and b-jet system (bottom
plots) in EW ME $gg \to t \bar t Z (\to b \bar b)$ and EW ME $gg \to (Z/W/\gamma^* \to)t \bar t b
\bar b$ events.  The distributions are plotted only for b-quarks (resp. b-jets) originating in
the hard process.
\label{FS4:c}} 
\end{figure}

Fig.~\ref{FS4:c} shows the respective invariant mass distributions of  the b-quark pair
(left plot) and the b-jet pair (middle and right plots) not originating from top-quark
decays in reconstructed $t \bar t b \bar b$ events, as estimated with either full or with
only resonant EW ME processes. The presence of the flat non-resonant component, which is
quite substantial with respect to the resonant one, is rather evident. One can also
clearly see how the shape of the EW background is smeared when going from b-quark distribution to
b-jet distribution.

\begin{figure}
\begin{center}
     \epsfig{file=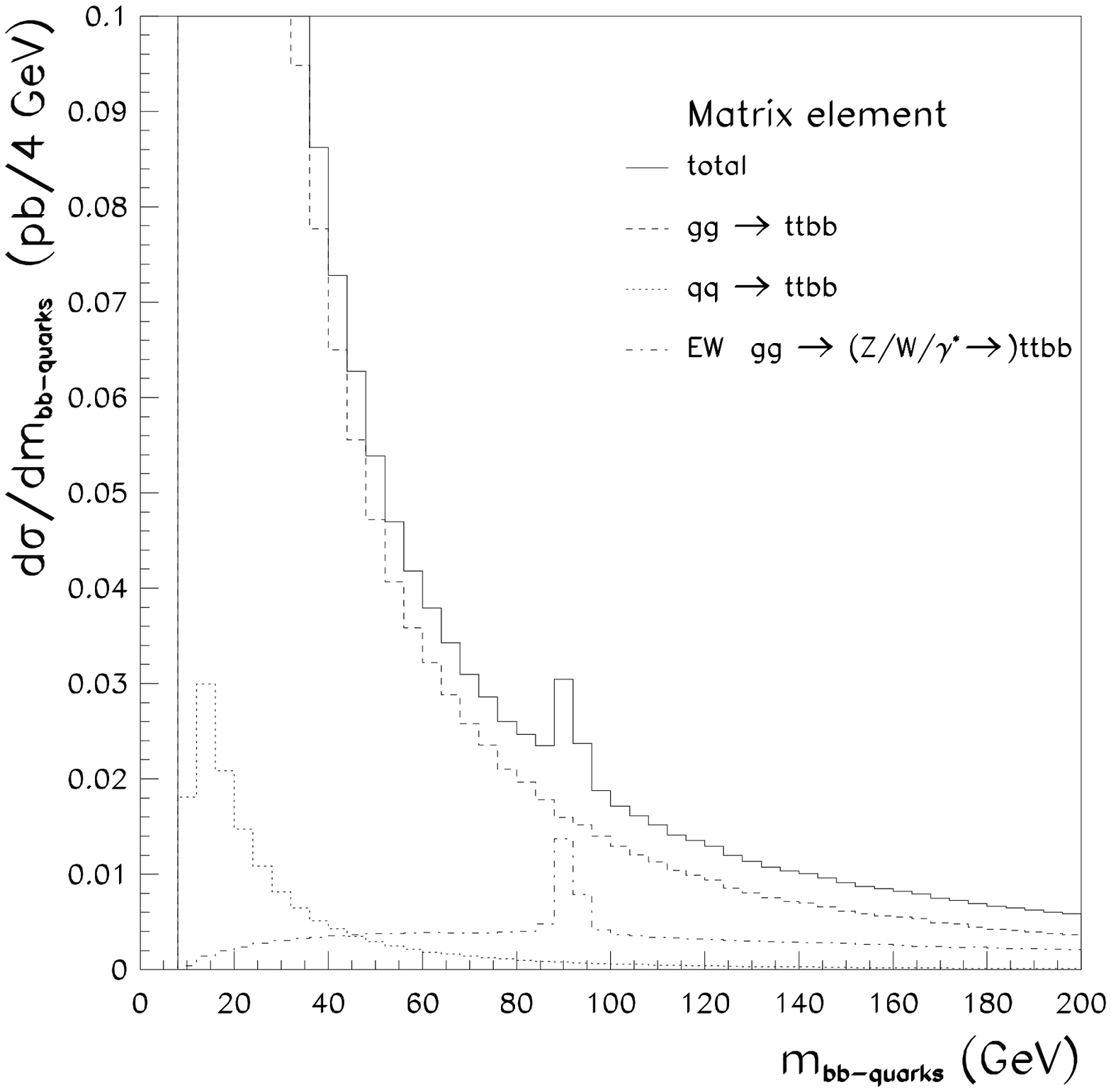,width=4.5cm}\\
     \epsfig{file=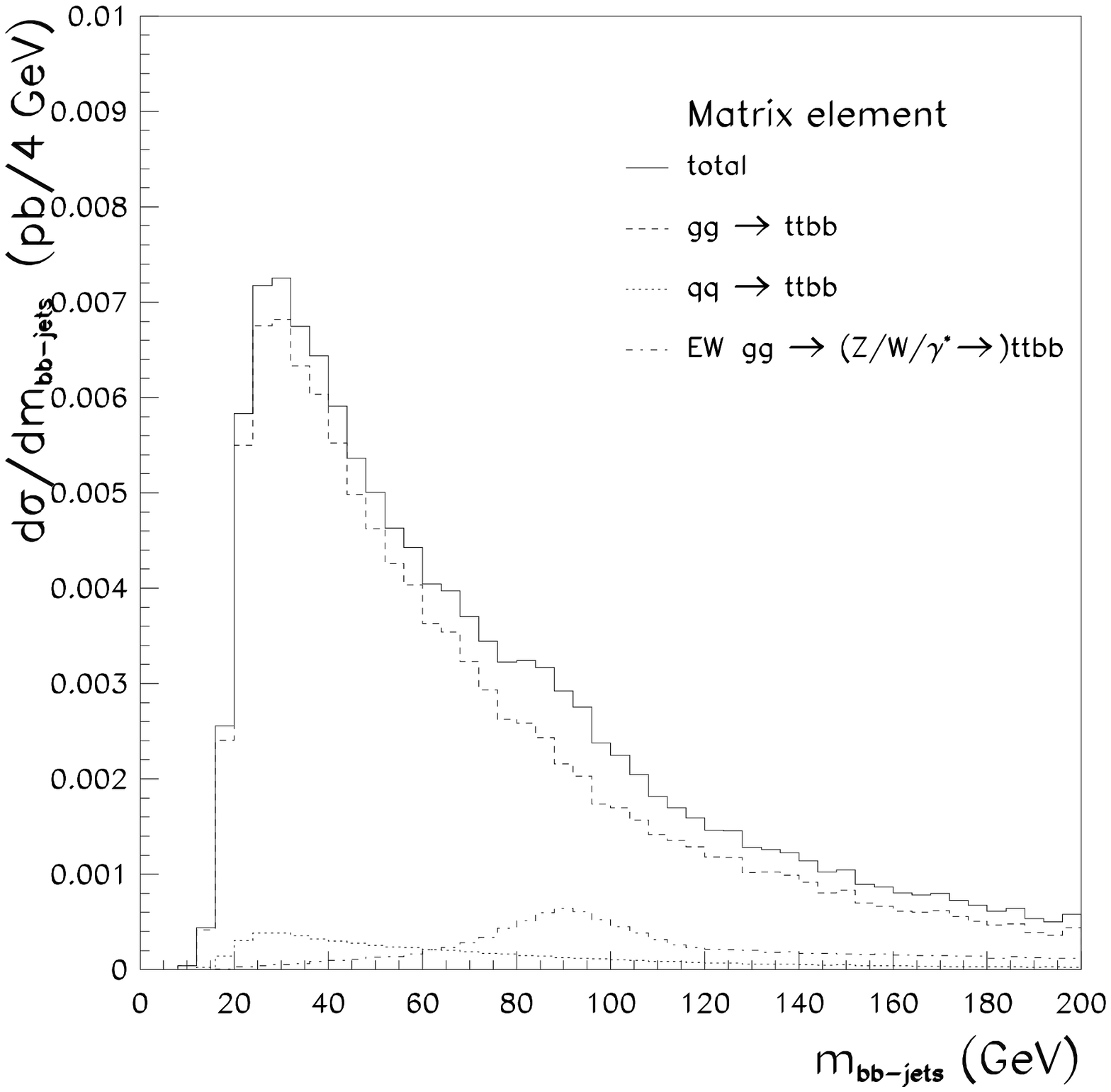,width=4.5cm\hskip -0.3cm}
     \epsfig{file=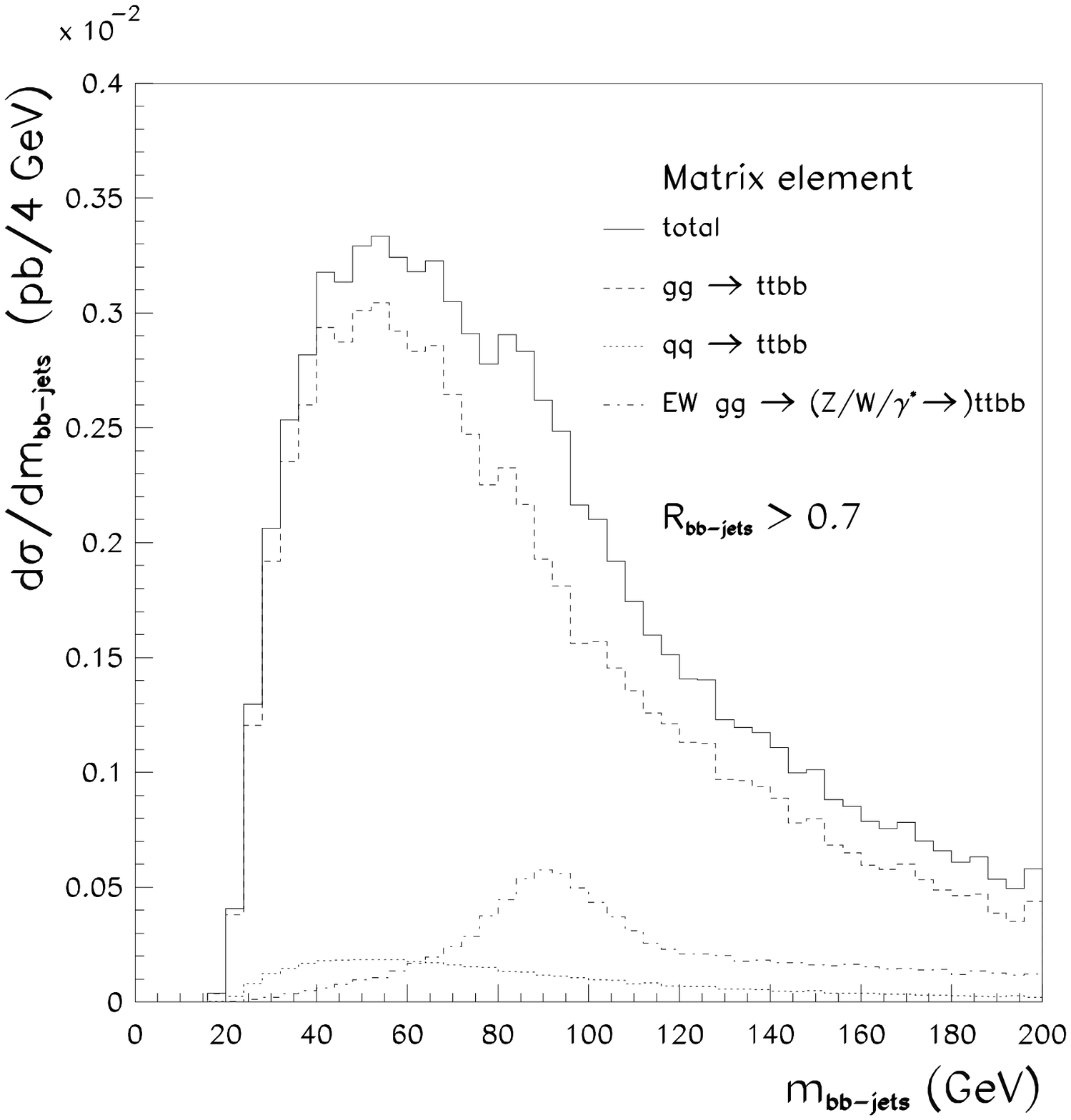,width=4.5cm}
\end{center}
\caption{\em
Invariant mass of the b-quarks system (top plot) and b-jet system (bottom plots)
in QCD ME and EW ME  events.
The distributions are plotted only for b-quarks (resp. b-jets)
originated from the hard process.
\label{FS4:d}} 
\end{figure}

Finally the QCD and EW component of the ME simulation chain are added
together. Fig.~\ref{FS4:d} again shows the respective invariant mass distributions of the
b-quark pair (top plot) and b-jet pair (bottom plots) in reconstructed $t \bar t b \bar b$
events. Also shown separately are the QCD $gg \to t \bar t b \bar b$, QCD $q \bar q
\to t \bar t b \bar b$ and EW $gg \to (Z/W/\gamma^* \to)t \bar t b \bar b$ components. The 
EW component contributes around 20-25\% of the QCD one in the mass range of interest.

It is crucial to notice that the EW component produces a resonant structure above the
non-resonant QCD+EW.  This will make a discovery of MSSM light Higgs in the $ t \bar t b
\bar b$ channel, with the mass between 95-110 GeV, even more difficult than assumed in
\cite{ActaB30}. One should however keep in mind that the relative contribution of  QCD and EW
components strongly depends on the chosen factorisation scale, see \cite{AcerMC}, and  will thus 
require very good theoretical understanding before the signal observability could be claimed at LHC
in this channel.

\section{ Conclusions}

In this paper a quantitative comparison between matrix element (ME) and
parton shower (PS) approaches for generating key irreducible backgrounds to the light
Higgs boson searches at LHC with  multi-b-jet final states was presented.  Three processes were discussed:
the $W b \bar b$, $Z b \bar b$ and $t \bar t b \bar b$ production. The {\tt AcerMC} Monte
Carlo generator was used for simulating ME events while {\tt PYTHIA 6.2} was used for
simulating PS events. 
\begin{itemize}
\item
For the $W b \bar b$ background there is a reasonable agreement of $p_T^W$ spectra in
$\ell b \bar b$ events. Expected rates are however almost 1.5-2.5 higher in PS simulation. This
enhancement comes mostly from configurations where b-jets are very close. Requiring large
cone separation between b-jets and vetoing events with additional jets brings the ME estimates 
to be in acceptable agreement for the relevant invariant mass range of the b-jet system. 
\item
For the $Z b \bar b$ background the PS approach is clearly underestimating the hardness
of the tail of the $p_T^Z$ spectra.  This is due to the
dominant contribution from the multipheral Feynman diagrams, not reproduced well even
with an {\it  improved parton shower} implementation in {\tt PYTHIA}. The distributions of
the invariant mass of the b-jet system and their normalisations are in reasonable agreement.
\item
There are two components of the $t \bar t b \bar b$ background, the QCD and EW ones. Only
the QCD component can be simulated with the PS approach. For QCD PS and
QCD ME events there  is a relatively
good agreement in the shape and normalisation of the invariant mass distribution of the
b-jet system. The transverse momenta distribution of
that system is however much harder in ME events. The EW ME component is not negligible and
leads to the resonant structure in the total QCD+EW background. This would make
observability of the mass range around the Z-boson mass more difficult than hoped so far.
\end{itemize}

It is quite evident that well understood theoretical predictions for these processes will
provide the key for establishing the Higgs signal observability; it is however difficult
to draw universal conclusions from the comparisons presented above. Therefore, it is
rather encouraging that the PS and LO ME predictions are not very far off. The 50\%
differences e.g. for $W b \bar b$ events in the overall normalisation are still within
expected uncertainties for this type of background estimates. It would nevertheless be
very important to perform a similar comparison with NLO ME predictions.  Such
implementations are already becoming available for $W b \bar b$ events \cite{NLOWbb} and 
$Z b \bar b$ events \cite{NLOZbb}. Nevertheless, at the time of the studies  
presented here, they were still not in the form allowing for straightforward applicability.

Given several possible tunings of the parton shower model as presently available in {\tt
PYTHIA}, one could probably easily improve further on the agreement between both
approaches.  The framework prepared in {\tt AcerMC} generator could well be a nice tool
for such a tedious task. It is however not clear, that the parameters should be tuned in a
way to make the PS predictions agree with the ME ones. Rather, an enhanced theoretical
understanding for which applications the PS or ME predictions are more credible and why
should be achieved first.  We hope that the results presented here could contribute to
such discussions.

Indeed, most of the ME/PS comparison studies so far concentrated predominantly on soft gluon
emissions and/or NLO effects, where the hard processes under study are very simple and/or
unproblematic (e.g. Drell-Yan Z boson production, $t \bar t$ production etc\ldots),
whereas this study was concentrated on the limiting case of processes involving either a
large set of diagrams (from ME viewpoint) or quite hard gluon emission (from PS
viewpoint). The latter case might be even more problematic since the soft (NLO) effects
have already been (to some extent) properly incorporated into the PS algorithms
\cite{Sjostrand1998,Sjostrand2000} whereas the hard gluon radiation (at high $p_T$) is a
feature that might be supposed not to work really well under PS approach by virtue of GLAP
equations \cite{GLAP} and has received relatively little study so far.  An additional
point to support this claim is that the fiducial kinematic cuts applied in the above
studies, which were chosen to reflect the cuts that will be applied in the future
experimental analyses, to a certain extent suppress the soft (NLO) effects (e.g. cuts on
minimal $p_T$, minimal cone separation etc..). This makes the discrepancies in the hard
process description the focal point of attention.

The issue of establishing a consistent procedure for the appropriate Monte Carlo
generation \cite{Field} is crucial as the {\it complete} evaluation of the expected
background should include its estimates for both irreducible and reducible\footnote{
'Reducible' events denote events where one or more jets without heavy-flavour content are
misidentified as b-jets} components \cite{ATL-PHYS-TDR}.  Quantifying the discrepancies
between the ME and PS simulation approaches is important because it indicates what could
be the expected systematic bias on the evaluation of the reducible backgrounds.  The fact
that PS and ME predictions are not very far off in the discussed cases of the irreducible
backgrounds is also encouraging for using PS approach for simulating their reducible
components.

Results from the presented stud\-ies thus give us a stron\-ger confidence in using the PS
approach for simulating the {\it complete} backgrounds. The ME approach alone would be not
sufficient but is nevertheless very valuable for quantifying the uncertainties of the PS
approach in case of the irreducible background components. The results however also
indicate that (for obvious reasons) the PS approach alone is also not without its
shortcomings. Some contributions, like e.g. the full EW $t \bar t b \bar b$ can presently
not be covered by the PS algorithms. One should thus definitely aim for having both
approaches available in a form which is straightforward to use in the experimental
analyses with a clear theoretical understanding of their individual shortcomings.


\begin{acknowledgement}
This work was performed within the framework of the Higgs Working Group of the ATLAS
Collaboration. We are grateful to all our colleagues for the inspiring atmosphere and
several very valuable discussions.  We have used a simplified version of the fast
simulation of the ATLAS detector for the quantitative evaluations of
the detector responses presented in this paper.
\end{acknowledgement}


\end{document}